\documentclass[preprint,review,12pt]{elsarticle}

\usepackage{epsfig}
\usepackage{graphicx}

\usepackage{amssymb}
\usepackage{amsthm}
\usepackage{amsmath}
\newcommand{\beq}{\begin{equation}}
\newcommand{\eeq}{\end{equation}}

\usepackage{geometry}

\usepackage{algpseudocode}

\usepackage{subcaption}

\usepackage{float}
\usepackage{multirow}

\journal{Journal of Computational Physics}

\begin{document}

\begin{frontmatter}



\title{
Robustness of complexity estimation in event-driven signals against accuracy of event detection method
%RTE-DSF: Rapid Transition Events Diffusion Scaling Finder Algorithm
}


\author[a1,a2]{Marco Cafiso\corref{cor1}}
\ead{marco.cafiso@phd.unipi.it}
\author[a2,a3]{Paolo Paradisi\corref{cor1}}
\ead{paolo.paradisi@cnr.it}
\cortext[cor1]{Corresponding author}
\affiliation[a1]{organization={Department of Physics 'E. Fermi', University of Pisa},%Department and Organization
            addressline={Largo Bruno Pontecorvo 3}, 
            city={Pisa},
            postcode={I-56127}, 
            state={},
            country={Italy}}
\affiliation[a2]{organization={Institute of Information Science and Technologies ‘A. Faedo’, ISTI-CNR},
             addressline={Via G. Moruzzi 1},
             city={Pisa},
             postcode={I-56124},
             state={},
             country={Italy}}
\affiliation[a3]{organization={BCAM-Basque Center for Applied Mathematics},
             addressline={Alameda de Mazarredo 14},
             city={Bilbao},
             postcode={E-48009},
             state={},
             country={Basque Country, Spain }}

\begin{abstract}
%The current wide availability of large datasets and of increased computing power enables 
%The current wide application of machine learning methods to large datasets has renewed the study of complex systems, which has gained new attention for its ability to extract structured and synthetic information from such data. 
%Complex systems are defined as systems with many interacting dynamic units, e.g., neurons or agents linked together in a complex network. 
Complexity has gained recent attention in machine learning for its ability to extract synthetic information from large datasets. 
Complex dynamical systems are characterized
%are not only characterized by topological properties, but also 
by temporal complexity associated with intermittent birth-death events of self-organizing behavior.
%, in which metastable states emerge and decay through rapid transitions. 
These rapid transition events (RTEs) can be modelled as a stochastic point process on the time axis, with inter-event times (IETs)
revealing rich dynamics. In particular, IETs with power-law distribution mark a
departure from the Poisson statistics and indicate the presence of nontrivial complexity that is quantified by the power-law exponent $\mu$ of the IET distribution. However, detection of RTEs in noisy signals remains a challenge, since false positives can obscure the statistical structure of the underlying process.
In this paper, we address the problem of quantifying the effect of the event detection tool on the accuracy of complexity estimation.
This is reached through a systematic evaluation of the Event-Driven Diffusion Scaling (EDDiS) algorithm, a tool exploiting event-driven diffusion to estimate temporal complexity.
After introducing the event detection method RTE-Finder (RTEF), we assess the performance of the RTEF-EDDiS pipeline using event-driven synthetic signals.
The reliability of the RTEF is found to strongly depend on parameters such as the percentile 
%or the time window used to evaluate the signal steepness 
and the number of false positives can be much higher than the number of genuine complex events. 
Despite this, we found that the complexity estimation is quite robust with respect to the rate of false positives. For the power-law distributed IETs with $\mu\le2.5$, the second moment scaling $H$ appears to even improve as the rate of false positives increases, reaching estimation errors of about $4-7\%$.
%Conversely, the distribution self-similarity $\delta$ has an opposite trend giving better results as the rate of false positives decreases
%
\end{abstract}

\begin{graphicalabstract}
\end{graphicalabstract}

\begin{highlights}
\item The estimation of temporal complexity is robust with respect to the presence of false positives in the event detection method.
%Research highlight 1
\item A counter-intuitive result is that, in the range of power exponent $\mu$ less than $2.5$ the accuracy in the evaluation of diffusion scaling $H$ improves as the number of false positives increases.
\end{highlights}

\begin{keyword}
Temporal complexity \sep intermittency \sep diffusion scaling \sep rapid transition events \sep power-law \sep noise in signal processing
\PACS 05.40.$-$a \sep 05.40.Fb \sep 05.45.Tp \sep 05.65.+b \sep 05.40.Ca
\end{keyword}

\end{frontmatter}


%%%%%%%%%%%%%%%%%%%%%%%%%% Introduction %%%%%%%%%%%%%%%%%%%%%%%%
\section{Introduction}
\label{sec:intro}

\noindent
The present-day focus on data science is linked, on the one hand, to the great availability of huge datasets and, on the other hand, to the increased computational power that makes it possible to use machine learning methods on these same datasets.
In this framework, complexity is living a renewed interest due to the need for extracting synthetic information from the large amount of available data \cite{niu-west_2021_fract,krakauer_fcs2023}.

\noindent
A complex system consists of many interacting elements, or nodes in a network, with nonlinear, cooperative dynamics that give rise to emergent, self-organized, structures, typically displaying some coherence among different elements and in time \cite{paradisi_csf15_preface}.
This coherence property can be quantified by proper {\it complexity} measures, 
or {\it metrics} in the information science jargon, 
%such as long-range correlations, 
computed from data that are obtained by observational probes embedded in the complex system.
Complexity measures are often referred to the topological properties of 
%these networks and of 
the emerging coherent structures and are investigated in the framework of complex networks \cite{boccaletti_pr2006,battiston_pr2020,ji-kurths_pr2023,artime_nrp2024}, with wide and interesting applications in the field of neuroscience \cite{bullmore_nrn2009,rubinov_ni2010_reviewcomplexnet,sporns_tcs2022_complexbrain,vaiana_jns2020,yeh_jmri2021,chiarion_be2023,borra_bsf2025,laasch_sr2025,gund_ans2025}.
%by means of the intricate link structure and many complexity studies are consequently focused on 
However, when dealing with time evolving systems, e.g., time-varying networks, each network node is represented by a temporal signal. 
In this case, the investigation focus is also on the temporal dynamics of the network and, thus, on the temporal evolution of self-organized emerging states \cite{grigolini_csf15_bio_temp_complex,paradisi_springer2017}.
%Actually, complexity in a multi-component systems is not only characterized by the topological features of emerging structures, but also on their temporal evolution.
Actually, the cooperative dynamics of a wide set of complex systems trigger the emergence of coherent, or self-organized, structures that are not equilibrium states of the system, but, on the contrary, display peculiar metastability characterized by random lifetimes
\cite{fingelkurts_ni2003,Kaplan2005RTPsAlgorithm,rabinovich_plos-cb2008,rabinovich_plr2012,fingelkurts_csf2013}. 
The complex metastability of these coherent states is a condition that is referred to as Temporal Complexity (TC) \cite{grigolini_csf15_bio_temp_complex,turalska-grigolini_pre11,grigolini_pre15,mahmoodi-grigolini_sr2024}  or Intermittency-Driven Complexity (IDC) \cite{paradisi_springer2017,paradisi_csf15_pandora}. In this case, the self-organized structures are not equilibrium states but are characterized by a lifetime after which the instabilities of the nonlinear interactions determine a rapid decay of the self-organized state.
After that, the system experiences a transition, typically a fast one, to a new dynamic condition, which may be another self-organized state or a disordered state.
The latter case is clearly defined by the absence of well-defined internal coherence, whereas, on the contrary, a self-organized state is represented by a state with high internal coherence, e.g. long memory and/or long-range correlations between nodes.
%In statistical data analysis the internal coherence of a state is evaluated with different approaches, e.g., by means of spatial and/or temporal covariance and phase locking.
%%%, which can be revealed, for example, through a spatial and/or temporal covariance in the ``observals''. 
The overall displayed behavior is that of a sequence of intermittent events marking the transition among different dynamical regimes.
%, that is, a intermittent birth-death process of self-organization.
%
It is essential to emphasize that transitions between self-organized states or between high and low coherence phases are often so fast as to be reduced to an instant in time.
In this framework we can introduce the concept of Rapid Transition Event (RTE), and the temporal sequence of RTEs can be
modelled as a stochastic point process in the time axis \cite{cox_1980_point-processes}, a concept that was largely exploited in brain studies \cite{fingelkurts_ni2003,Kaplan2005RTPsAlgorithm,allegrini_pre2009,allegrini_pre10,tagliazucchi-chialvo_fp2012,tagliazucchi-chialvo_fn2016}.
A point process is mathematically represented as a set of strictly increasing time instants $\{ t_n ; n \in \mathcal{N}\}$, being $\mathbf{N}$ the set of integer numbers $n \ge 0$.
%, labeling a succession of crucial events, herein referred to as Rapid Transition Events (RTEs).
%
The most important features of the intermittent birth-death process of self-organizing behavior are then the Probability Density Function (PDF) $\psi(\tau)$ computed from the sequence of Inter-Event Times (IETs) $\tau_n = t_n - t_{n-1}, n\ge 1$,
and the statistical dependence among RTEs and, consequently, among IETs.
The {\it renewal} condition, which is defined by the mutual independence of IETs in the point process,
is a further important feature that often characterizes RTEs \cite{cox_1970_renewal}. %A renewal point process is defined by the mutual independence of IETs between two consecutive RTEs 
The most interesting examples of IET-PDFs are given by an exponential, indicating a Poisson process, stretched exponential and power-law decays:
$$
\psi(\tau) \sim e^{-r_p \tau};\ e^{-\left(r_p \tau\right)^\alpha};\ \frac{1}{\tau^\mu}
$$
with $0 < \alpha < 1$ and $1 < \mu < 3$.
While the Poisson process is associated with the lack of self-organization, 
the last two behaviors, in particular the power-law, identify important non-Poisson processes marking the intermittent emergence of self-organized states \cite{paradisi_springer2017}.

\noindent
In IDC systems, the testing of the renewal condition and the IET-PDF shape can be exploited to define the level of complexity \cite{paradisi_springer2017,allegrini_pre2009}. Inverse power-law decay in the IET-PDF represents the most interesting case, as it links to self-similarity spanning over a wide range of time scales.
In this case, the power exponent $\mu$ is defined as the IDC index, thus being associated with the capacity of the cooperative dynamics to trigger self-
organized metastable states
\cite{paradisi_springer2017}.
Some prototypical models of signals with complex intermittency 
are given by the Pomeau-Manneville map, where the presence of a marginally stable point determines a sequence of fast transitions between a laminar and a chaotic region, \cite{manneville_1980, pomeau-manneville_1980,allegrini_pre2003_model},
and by heteroclinic channels
%in dynamical system theory 
\cite{rabinovich_plos-cb2008,rabinovich_plr2012,rabinovich_prl2006}, that is, a sequence of fast passages through several saddle points that define the fast transitions between two different regions of the state space.

\noindent
Event-based or intermittency-based approaches, i.e., based on the detection and statistical characterization of RTEs, were extensively applied in several fields. 
A greatly interesting case is given by brain dynamics, which are probably the most important prototype of a complex system \cite{fingelkurts_csf2013,bullmore_nrn2009,rubinov_ni2010_reviewcomplexnet,sporns_tcs2022_complexbrain}. 
%Similar complex bevahiour is also found in {\it in vitro} neural networks \cite{}.
% Qui andrebbe forse accennato alla Self-Organized Criticality \cite{bak,chialvo, dearcangelis, ora anche jiirsa e sorrentino e scarpetta}
In brain studies, the authors of Refs. \cite{fingelkurts_ni2003,Kaplan2005RTPsAlgorithm} firstly introduced the idea of RTE\footnote{
%%%%%%%%%%
These authors actually named Rapid Transition Processes (RTPs) the here introduced concept of RTEs.
%%%%%%%%%%
}
in the neuroscience field and applied to the processing of ElectroEncephaloGrams (EEGs) and MagnetoEncephaloGrams (MEG). 
Other authors applied similar event-based approaches and similar RTE definitions
to investigate brain complexity 
%in basal condition 
\cite{allegrini_pre2009,allegrini_pre10} 
%and during sleep \cite{allegrini_pre15}, 
%%Similar approaches were later applied to 
funtional Magnetic Resonance Imaging (fMRI) data \cite{tagliazucchi-chialvo_fp2012,tagliazucchi-chialvo_fn2016}, neural dynamics \cite{turalska-grigolini_pre11,zare-grigolini_csf2013,cafiso_proceed-ncta2024}.
%TC/IDC approach to complexity was successfully applied to different systems, including ???????
The concept of RTE  was also applied in several other research contexts and to different kinds of data, e.g., atmospheric turbulence \cite{paradisi_npg12}.
%and atmospheric pollution \cite{paradisi_epjst09}.
%
In general, many different RTE definitions were exploited including, more in general, the definition of crucial transition events not directly linked to rapid passages but, e.g., to some threshold passage in the signal, often after proper processing of the signal itself \cite{paradisi_npg12}.

\noindent
The event-based approach is a basic building block of the TC/IDC framework, such as the RTE definition and the associated event detection algorithm. 
%In particular, sequences of RTE represent the starting point of the EDDiS\footnote{} algorithm
In this respect, the effect of noise and the presence of false positives in the event detection are unavoidable.
However, in complex systems theory, the main interest lies in the reliable estimation of the system's complexity. 
Regarding the IDC class of complex systems,
%, i.e., systems characterized by rapid transitions among metastable states, 
the proper evaluation of complexity indices can be affected by the goodness of the event detection algorithm, whose accuracy can be affected by noise and by the unavoidable presence of false positives. 
In particular, the presence of noisy, secondary, events has a blurring effect on the WT-PDF \cite{allegrini_pre10,paradisi_csf15_pandora}, thereby foreclosing the possibility of a reliable estimate of the IDC index $\mu$ and, in general, or other properties related to the IET-PDF and to event sequence.
However, 
%regarding the estimation of complexity features in a temporal signal, 
a method to treat with noisy or secondary events is the Event Driven Diffusion Scaling (EDDiS) algorithm \cite{paradisi_springer2017,paradisi_csf15_pandora,allegrini_pre2009}, a method that is based on the concept of anomalous diffusion and on related well-known results found in classical literature \cite{grigolini2001asymmetric,klafter_1987_ctrw_anom_diff,montroll1964random,montroll_etal-jmp-1965,montroll_1969_3,montroll_1973_4,scafetta_2002_dea,shlesinger_1974_ctrw_asympt,tunaley_jsp1974_ctrw_asympt,tunaley_1975_ctrw_asympt,tunaley_1976_ctrw}. 
This algorithm is based on the idea that secondary events drive a normal diffusion process while crucial complex events trigger anomalous diffusion \cite{paradisi_springer2017,metzler_pr2000,grigolini2001asymmetric,akin_pa06,akin_jsmte09}, this last one being defined by a non-Gaussian distribution and/or a nonlinear growth of the variance\footnote{
%%%%%%%%%
Here $X(t)$ is the integral of an observed signal $S(t)$, i.e., $\dot X (t) = S(t)$.
%%%%%%%%%
}:
\beq
\langle X(t)^2 \rangle \sim t^{2H} \ .
\nonumber
\eeq
$H$ is the second-moment scaling, which corresponds to the Hurst exponent for mono-scaling signals.
EDDiS algorithm is based on well-known theoretical results about diffusion processes driven by crucial events, in particular referring to the Continuous Time Random Walk (CTRW) model
\cite{metzler_pr2000,montroll1964random,montroll_etal-jmp-1965,weiss1983random,shlesinger_1987_ctrw_turb,zaburdaev-klafter_2015_levywalk}.
In particular, these theoretical results follow from the assumption of the renewal condition introduced above \cite{cox_1970_renewal}.

\vspace{.1cm}
\noindent
Even if the problem of noisy events in complexity evaluation was identified and the potentiality of EDDiS was investigated \cite{paradisi_csf15_pandora,allegrini_pre2009,allegrini_pre10}, a systematic evaluation of the accuracy in the complexity estimation is
still lacking.

\vspace{1.cm}
In this paper, we carry out a systematic study of estimation errors in complexity indices derived from the EDDiS algorithm. In particular, we investigate the role of the event detection algorithm and characterize the robustness of complexity estimation concerning the choice of parameters that need to be fixed {\it a priori} in the event detection algorithm.
To carry out this study we here propose and develop an extension of an event detection algorithm first proposed in Ref. \citet{Kaplan2005RTPsAlgorithm}. Our approach
is essentially a generalization of that algorithm and of later variations proposed and applied by \citet{allegrini_pre2009,allegrini_pre15}. 
The proposed algorithm of event detection is then applied to a time series generated by a stochastic model that combines a sequence of complex events with a damped oscillator and an additive white Gaussian noise component.

\vspace{.1cm}
\noindent
The paper is organized as follows.
In Section \ref{sec:methods} we introduce the event detection algorithm, hereafter referred to as RTE Finder (RTEF), we briefly recall the EDDiS tool and we define the metrics used to evaluate the estimation errors.
In Section \ref{sec:Signal_Generator} we introduce our stochastic model used to generate
artificial signal driven by events with IET distributed according to an inverse power-law with a power exponent $\mu$ (complex events) or to an exponential (Poisson events). In Section \ref{sec:num_sim} we illustrate the results of numerical simulations and, in particular, the application of the RTEF-EDDiS pipeline of statistical analyses to the simulated event-driven signals. In subsection \ref{sec:discussion} we give a detailed discussion of the results. Finally, in Section \ref{sec:conclusions} we discuss our findings.

%%%%%%%%%%%%%%%%%%%%%%%%%%%%%%%%%%%%%%%%%%%%%
%%%%%%%%%%% Methods of Data Analysis %%%%%%%%%%%%%%%%%%%%%%%%%%%%%%%%%%%%%%%%%%%%%
\newpage
\section{Methods of Data Analysis}
\label{sec:methods}
\noindent
Our goal is to test the accuracy in the estimation of IDC index $\mu$ by the combined applications of an event detection algorithm, herein denoted as RTE-Finder (RTEF), and of the EDDiS algorithm.

% Inspired by the work of Kaplan et al. \citet{Kaplan2005RTPsAlgorithm}, we implement a novel algorithm for Rapid Transition Processes (RTPs) detection. In particular, we develop an algorithm to find the probability distribution of the RTPs and we test it with synthetic data.

\subsection{The RTEF algorithm}
% Vecchio RTE-DSF = RTE - Diffusion Scaling Finder
\label{sec:RTEF}

\noindent
Our proposed algorithm is inspired by and generalizes the work of  \citet{Kaplan2005RTPsAlgorithm} and of \citet{allegrini_pre2009,allegrini_pre15}.
The main differences lie in the evaluation of the signal steepness,
%by means of a derivative estimation, 
of the signal envelope and on the possibility of extending the analysis to different signal frequency bands.

%Inspired by the work of \citet{Kaplan2005RTPsAlgorithm}, 

\noindent
The RTEF algorithm consists of two main steps: (1) calculation of the signal envelope and (2) 
extraction of the RTEs. In the first step, we estimate the signal envelope by using the modulus of the analytical signal associated with the 
Hilbert transform $\mathcal{H}$, i.e.:
$$
S_a(t) = S(t) + i\ \mathcal{H}[S](t)
$$
so that the envelope is given by $E(t) = |S_a(t)|$
Following the convention by \citet{Kaplan2005RTPsAlgorithm}, the signal envelope $E(t)$ is hereafter denoted as ``Test Signal''.
% RTE-finder algorithm
% is performed in two main steps: (1) Detection of preliminary RTEs; and (2) Selection of the true RTEs from those found at point (1). 
% In the first step, we estimate the signal's envelope by using the absolute value of the Hilbert transform. From the Hilbert transform of the signal, we obtain two signals: (a) a Test Signal that is the absolute value of the Hilbert transform; (b) a Level Signal that is a sequence derived by applying the moving average, with a relatively large window (usually around 200-300 time-steps), on the Test Signal obtained at point (a). Finally, we detected what we called 'preliminary RTEs' by extracting the time location of the intersection of the signals obtained at points (a) and (b).
In the second step, we start from the Test Signal and calculate the approximation of its derivative to see the rapid changes in the signal. The derivative of the Test Signal at time $t_k$ is estimated by the following formula: %with the classical formula of the numerical differentiation:
\begin{equation}
    \label{derivative_approx}
    \dot E(t_k) = \frac{1}{N_d} \sum_{j = 1}^{N_d} \frac{E(t_k + j T_s) - E(t_k - j T_s)}{2 \cdot j \cdot T_s}
\end{equation}
being \(N_d\) the time window that we consider to estimate the derivative and $T_s$ the sampling time of the signal. Eq. (\ref{derivative_approx}) is an average of \(N_d\) difference quotients, centered at $t_k$, taken at different time intervals $2 \cdot j \cdot T_s$ around $t_k$ itself. 
For the estimation of the derivative we considered two time windows: \(N_d = 5\) and \(N_d = 12\). We only report here the results obtained with \(N_d = 5\), while the analyses carried out with $N_d=12$ can be found in the Supplementary Material.

\noindent
Finally, we compute the absolute value of the derivative and we determine a threshold by selecting a specific percentile value from the distribution of these absolute values. 
The RTEs are then defined by the threshold crossings of the absolute value of the derivative. 
%Since we focus only the positive RTEs, we took only the positive values of the thresholded estimated derivative.
%%%%%%%%
%% Finally, we define the so-called 'Real RTEs' as the intersection between the 'preliminary RTEs' defined at point (1) and the RTEs detected by taking a percentile from the derivative distribution. % The algorithm is summarize below:
% \begin{algorithmic}[1]
% \State $i \gets 10$
% \If{$i\geq 5$} 
%     \State $i \gets i-1$
% \Else
%     \If{$i\leq 3$}
%         \State $i \gets i+2$
%     \EndIf
% \EndIf 
% \end{algorithmic}
%%%%%%%%%%%%%%
% \subsection{Statistical Analysis}
% Since our objective is to see if the distribution of WTs of detected RTEs concerning the real distribution of WTs are similar, we decided to use two types of statistical tests: (1) the Mann-Whitney test; and (2) the Kolmogorov-Smirnov test. Moreover, we used also a tool from the theory of Intermittency-Driven Complexity, also known as Temporal Complexity, that is the Detrended Fluctuation Analysis (DFA) on the real and detected RTEs. This last method will be explained in detail in the next section. \\ 
% All the codes for signal simulations, RTEs distribution finder, and analysis are written in Python language and available at: (??????????????? Insert GitHub link).

\subsection{EDDiS Algorithm}
\label{sec:EDDiS}

\noindent
The EDDiS method is
a combination of two scaling analyses applied to three
different random walks driven by a sequence of events.
We here limit to the application of one particular random walk, based on the Asymmetric Jump (AJ) rule, and recall the 
two scaling analyses carried out on the resulting diffusion process.
Further details of the EDDiS algorithm can be found in Refs. 
\cite{paradisi_springer2017,paradisi_csf15_pandora,allegrini_pre2009} and, for reader's convenience, in the Supplementary Material.

\noindent
The two scaling analyses are given by the Diffusion Entropy \cite{grigolini2001asymmetric,akin_pa06,akin_jsmte09}, which evaluate the similarity $\delta$ of the diffusion PDF, 
and by the Detrended Fluctuation Analysis (DFA) \cite{peng_pre94}, estimating the second moment scaling $H$, essentially corresponding to the so-called Hurst exponent \cite{hurst_1951}.
The scaling exponent $H$ of the second moment is defined as:
\begin{equation}
    \label{h_scaling}
    \sigma^2(t) = \langle \left(X(t) - \overline{X}(t) \right )^2 \rangle \sim t^{2H}
\end{equation}
being $X(t)$ the diffusion variable resulted after applying the walking rule, and $\overline{X}(t)$ is a proper local trend of $X(t)$. 
The self-similarity index $\delta$ is defined as:
\begin{equation}
    \label{delta_scaling}
    P(x,t) = \frac{1}{t^{\delta}}F\Big( \frac{x}{t^{\delta}}\Big)
\end{equation}
being $P(x,t)$ the PDF of the diffusion variable $X(t)$.

\noindent
In general, the diffusion variable is written as the integral of a temporal signal:
$$
X(t) = \int_0^t \xi(t^\prime) dt^\prime \approx \sum_{n=0}^{N} \xi_n 
$$
being $t = t_N$.
In the AJ walking rule here applied, the impulsive, time-sampled, variable $\xi_n$ consists of making a unitary jump ahead when an event occurs. 
The resulting event-driven diffusion process
built according to the AJ rule 
corresponds to the counting process generated by the event sequence:
\begin{equation}
    \label{Asymmetric_Jump}
    X(t) = \#\{n : t_n < t\}.
\end{equation}
The functional relationships $\delta = \delta(\mu)$ and $H = H(\mu)$ in the case of AJ are the following \cite{paradisi_springer2017, grigolini2001asymmetric}:
\begin{eqnarray}
    &H_{AJ} = \begin{cases}
        \frac{\mu}{2} & \text{if } 1 < \mu < 2\\
        2 - \frac{\mu}{2} & \text{if } 2 \leq \mu < 3\\
        \frac{1}{2} & \text{if } \mu \geq 3
    \end{cases}
    \label{H_AJ}\\
    \ \nonumber \\
    & \delta_{AJ} = \begin{cases}
        \mu - 1 & \text{if } 1 < \mu < 2\\
        \frac{1}{\mu - 1} & \text{if } 2 \leq \mu < 3\\
        \frac{1}{2} & \text{if } \mu \geq 3
    \end{cases}
    \label{delta_AJ}
\end{eqnarray}
For this study, we apply this method directly to the detected RTE sequences.
This type of analysis is a powerful tool for scaling detection and, under the assumption of superdiffusion, i.e., $H>0.5$
and of sufficiently long sequence of events,
%, when applied to a sequence of transition events, 
is effective in detecting the anomalous diffusion scaling, thus giving 
useful information on the underlying dynamics that indeed generate the events. 
In the following, $H$, $\delta$ and $\mu$ will be generically referred to as {\it complexity indices}.

\subsection{Error estimates}
\label{sec:error}

To have a metric of performances of the proposed algorithm, we calculate the relative error for different metrics:
\begin{equation}
    \label{relative_error}
    RE = \Big|\frac{EV - RV}{RV}\Big|
\end{equation}
where $EV$ is the estimated value, and $RV$ is the real value.
When the $RE$ is referred to a set of numerical simulations, we compute the average of the RE over these same simulations and denote it as Mean RE ($MRE$).

%%%%%%%%%%%%%%%%%%%%%%%%%%%%%%%%%%%%%%%%%%%%%
%%%%%%%%%%%%%%%%%%%%%%%%%%%%%%%%%%%%%%%%%%%%%
%%%%%% Time Series Generator Model %%%%%%%%%%
%%%%%%%%%%%%%%%%%%%%%%%%%%%%%%%%%%%%%%%%%%%%%
\section{The time series model}
\label{sec:Signal_Generator}

\noindent
The model used to simulate the time series, or temporal signals, is inspired to the work of \citet{Guardabasso1988SignalSimulator}.
In particular, Guardabasso and co-workers studied a noisy relaxation process triggered by a sequence of random events with Poisson statistics.
We introduce here an extension of their model by introducing two elements:
\begin{itemize}
\item 
A Non-Poisson distribution of events according to a IET-PDF with inverse power-law decay\footnote{
%%%%%%%%%%%
The analysis of complex events, associated with an inverse power-law IET-PDF, is here compared with the standard Poisson events investigated by \citet{Guardabasso1988SignalSimulator}.
%%%%%%%%%%%
}.
\item
%noisy
A damped oscillator that is triggered by each event and whose dynamics are modeled through a second-order differential equation. 
\end{itemize}
Similarly to \citet{Guardabasso1988SignalSimulator}, a Gaussian white noise component is then added to the synthetic time series.

\noindent
In summary, the time series model is divided into four main steps: 
%%%%%%%%%%%%%%%%
\begin{itemize}
\item[(a)]
Generation of an event sequence, described by the corresponding set of event occurrence times: $\{ t_n; 1\le n \le M \}$.
\item[(b)] 
Generation of a pulse amplitude $A_n$ for each event, with $A_n$ uniform random number in $[1,10]$. At this point, we get a train of pulses with different amplitudes and the point process is described by the set $\{(t_n,A_n); 1\le n \le M\}$ and by the formula:
\begin{equation}       
S_{_{PT}}(t) = \sum_{i = 0}^{M} A_i \delta(t - t_i)
\label{pulse_train}
\end{equation}
\item[(c)] 
Convolution with pulse response: for each event, this step generates a damped oscillation, which then generalizes the relaxation process of \citet{Guardabasso1988SignalSimulator}.
\item[(d)] 
Addition of Gaussian white noise to the generated time series.
\end{itemize}
In step (a) a sequence of IETs, also known as Waiting Times (WTs), is first drawn from a
given IET-PDF. In particular, we choose two types of probability distributions \cite{paradisi_springer2017}:
\begin{itemize}
    \item Inverse power-law distribution for complex events \cite{paradisi_csf15_pandora,grigolini2001asymmetric,akin_jsmte09}:
\begin{equation}
\psi_n(\tau) = \frac{\mu - 1}{T} \frac{1}{(T+\tau)^{^\mu}}
\label{powerlaw_wt}
\end{equation}
being $\mu$ the power-law exponent and $T$ the time scale at which the power law emerges\footnote{
%%%%%%%%%%
When $\mu > 2$, the mean IET is given by $\langle \tau_n \rangle = T/(\mu - 2)$.
%%%%%%%%%%
}.
\item Exponential distribution for Poisson events
\begin{equation}
\psi_p(\tau) = r_{_p} e^{- r_{_p} \tau}
\label{exp_wt}
\end{equation}
where $r_p$ is the Poisson's event rate and $\langle \tau_p \rangle = 1/r_p$.
\end{itemize}
For both cases we set the mean IET equal to $1$ ($\langle \tau_p \rangle=\langle \tau_n \rangle=1$) and we draw a sample of $M = 20000$ IETs to get a sequence of random events.
In the case of power-law distribution, we draw the IETs by using the following formula \cite{allegrini_pre03}:
\begin{equation}
    \label{PL_distribution}
    \tau_i = T \left(\xi_i^{\frac{1}{1-\mu}} - 1 \right)\ ;\ i=0,1,...
\end{equation}
being $\xi_i$ a random number uniformly distributed in $[0, 1]$\footnote{
%%%%%
This formula is also obtained by applying the cumulative function method. 
%%%%%
}.
%The equation \eqref{PL_distribution} is taken from Allegrini's stochastic model \cite{Allegrini_pre03}. 

\begin{figure}[H]
    \centering
    \includegraphics[scale=0.5]{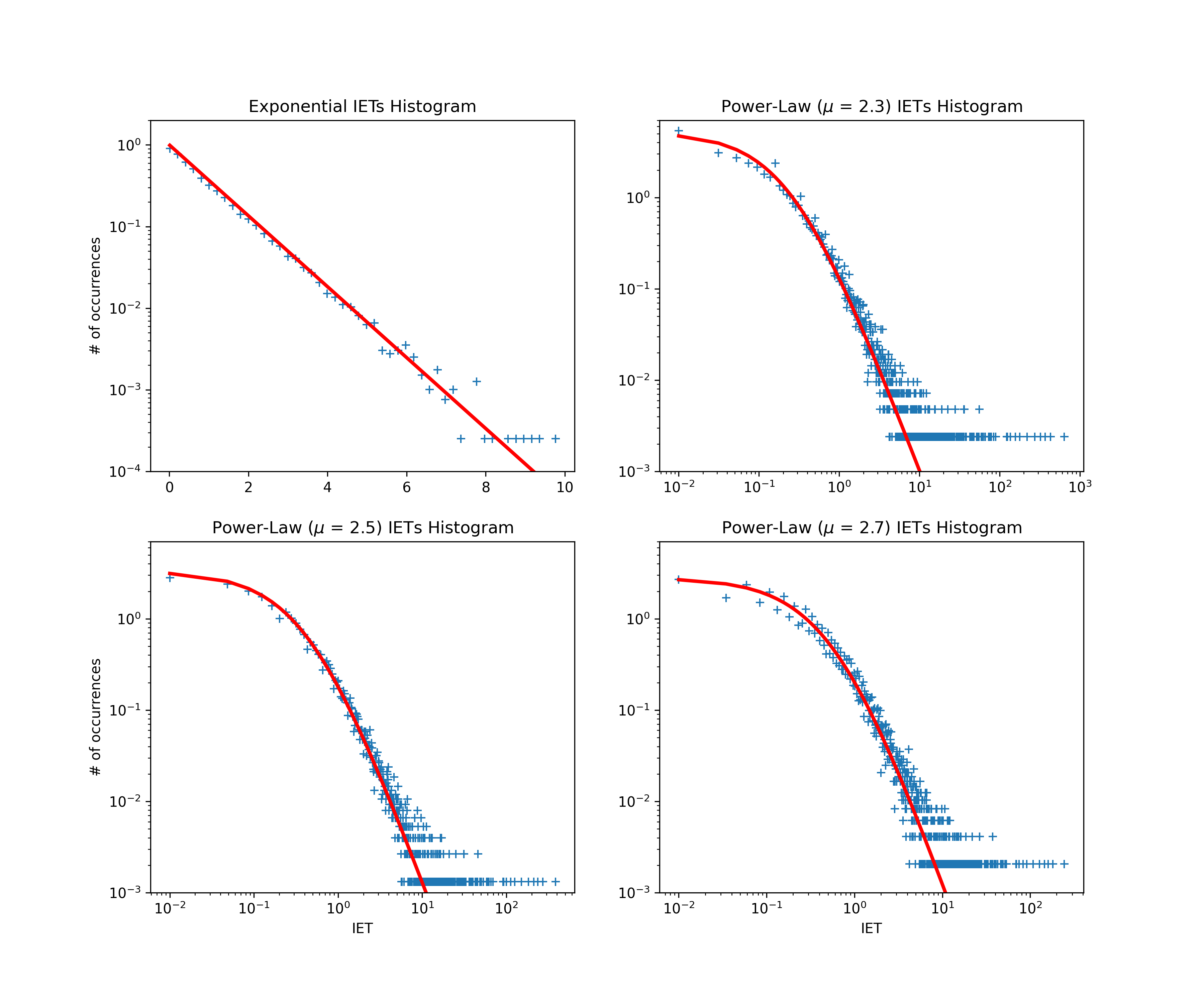}
    \caption{Comparison of observed IET distributions with theoretical curves.}
\label{fig:ExTimeSeriesWTsDistributions}
\end{figure}

\begin{figure}[H]
    \centering
    \includegraphics[scale=0.5]{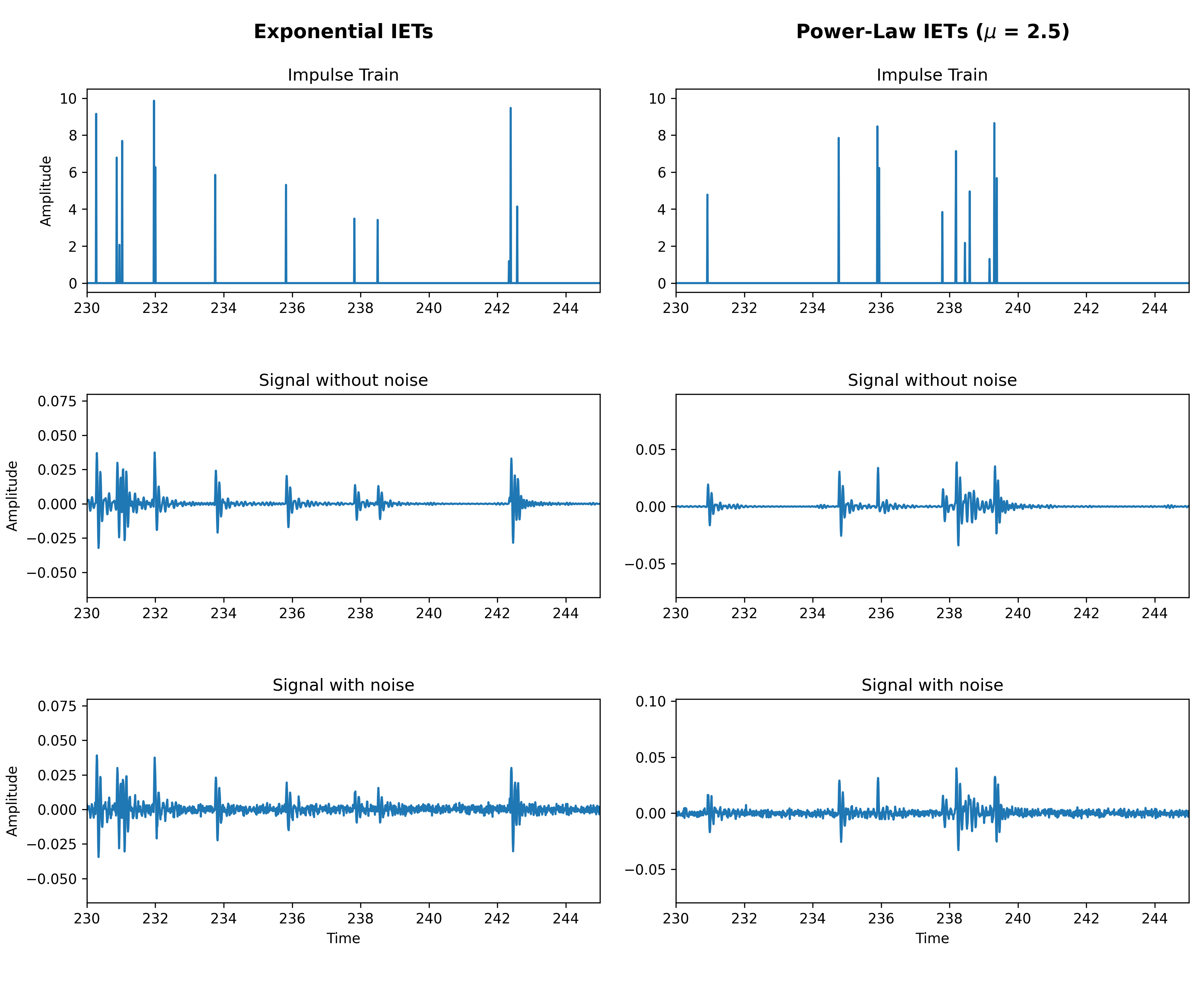}
    \caption{Example of time series generated by our model. Left and right panel refer to exponential and power-law (with $\mu = 2.5$) IET distributions, respectively. Top: comparison of the impulse trains. Middle: time series generated without the noise in frequency band $[8, 12] Hz$. Bottom: final time series with noise (with $10\%$ variance of the deterministic component).}
    \label{fig:ExTimeSeriesGenerator}
\end{figure}

\noindent
The IETs from the exponential distribution are drawn using the usual expression:
\begin{equation}
    \label{Exp_distribution}
    \tau_i = -\frac{1}{r_p}\log{\xi_i}\ ;\ i=0,1,...
\end{equation}
where, as in Eq.\eqref{PL_distribution}, $\xi_i$ is a random number uniformly distributed in $[0, 1]$. In Figure \ref{fig:ExTimeSeriesWTsDistributions} some generated IET-PDFs are reported together with the related theoretical curves, making it evident that the qualitative comparison is very good. 
Then, the sequence of event occurrence times is simply defined by:
$$
t_{n} = t_{n-1} + \tau_n\ ;\ t_0 = 0\ ; \ n=1,2,...,M
$$
The total simulation time is given by \(t_{max} = t_M\). In order to reproduce a sampled signal,
%Following the IET extraction, 
we then established a sampling grid spanning from 0 to \(t_{max}\) with a sampling time $T_s = 0.01$\footnote{
%%%%%%
When considering time units in seconds, this corresponds to a sampling rate of $100$Hz.
%%%%%%
}.
%%%time-step interval. 
%Here, \( t_{max} \) represents the summation of all extracted IETs, i.e., the total simulation time. 
Starting from Eq. (\ref{pulse_train}), the discrete time version of the pulse train is given as a sequence of zeros at each time step except for the steps that are nearer to an event occurrence time. In formulas, given $t_i = i \cdot T_s$ with $t_i$ the time step that is nearest to the event occurrence time $t_n$, it results: 
$S_{_{PT}}(t_i) = A_n$.
%
%Subsequently, we build a pulse train by aligning pulses $A_n$ with the endpoints of each IET within this grid, i.e., $t_l approx \t_n$. %%These pulses were assigned random amplitudes uniformly distributed between $1$ and $10$.
%
%% We decided to get random instead of fixed amplitudes to get an amplitude heterogeneity that can better mimic the fluctuations of a real time series.
%
Subsequently, according to above point (c), we generated a time series based on the pulse train using the following algorithm:
%%%%%%%%%%%%%%%%%%
\begin{enumerate}
\item 
    Take an angular frequency band: $[\omega_{0_{min}}, \omega_{0_{max}}]$, or equivalently, a frequency band: $[\nu_{min}, \nu_{max}]$, being that $\omega = 2\pi \nu$. 
\item 
    Choose $N$ angular frequencies $\omega_{0_{k}}$ equispaced in the band. In our case the grid size $\Delta \omega_0 = 1$.
\item 
    For each $\omega_{0_{k}}$ the pulse response function is given by a damped oscillator:
    \begin{equation}
    \label{kth_response_function}
    I_k(t) = \frac{e^{-\frac{\gamma}{2}t}}{\omega_{1_{k}}} \sin{\left( \omega_{1_{k}} t \right)}    
    \end{equation}
    where $\omega_{1_{k}} = \frac{1}{2}\sqrt{4\omega_{0_{k}}^2 - \gamma^2}$, and $\gamma$ is the relaxation factor of the damped oscillator. Notice that this is the solution of a second-order differential equation representing the dynamics of a particle in a harmonic potential with friction, i.e.:
    \begin{equation}
    \frac{dI_k}{dt} = v_k\ ;\quad \frac{dv_k}{dt} = - \gamma v_k - \omega_{0_{k}}^{2} x_k + \delta(t)
    \label{damped_oscill_eq}
\end{equation}
   being $I_k(0) = 0$, $v_k(0) = 1$\footnote{
%%%%%%%
The response function (\ref{kth_response_function}) is also referred to as a second-order linear filter.
%%%%%%%
    }.
The Dirac function $\delta(t)$ in Eq. (\ref{damped_oscill_eq}) is a pulse at time $t=0$ and the solution to this equation, given in Eq. (\ref{kth_response_function}), is the fundamental solution (Green function) of the system of differential equations in Eq. (\ref{damped_oscill_eq}) and, thus,  
the response to this initial pulse. 
\item
Then, for each event occurrence time $t_n$, the total pulse response function is given by the superposition of all contributions in the frequency band:
    \begin{equation}
    \label{total_response_function}
        I(t) = \sum_{k=1}^{N} I_k(t) = \sum_{k=1}^{N} \frac{e^{-\frac{\gamma}{2}t}}{\omega_{1_{k}}} \sin{\left( \omega_{1_{k}} t \right)}
    \end{equation}
%For each frequency in the band, the actual response is given by substituting $\delta(t)$ with A   
%
\item     
The total generated time series is then given by the time convolution between $I(t)$ and $ S_{_{PT}}(t)$:
$$
S(t) = (I*S_{_{PT}})(t)
$$
The sampled signal $S(T_i) = S(i)$\footnote{
%%%%%%%%
Since there are no ambiguities, it is possible to use the concise notation $S(i)$ for the sample signal.
%%%%%%%%
} 
is
%The convolution between $I(t)$ and $S_{_{PT}}(t)$ was 
calculated by Z-transforming the function \(I(t)\) and by multiplying it with the Z-transform of the pulse train $S_{_{PT}}(t)$. At the end of this step, given the sampling time $T_s$ and the time steps $t_i = i \cdot T_s$, we get a sampled signal $S(i)$ by anti-transforming.
     %\footnote{
%%%%%%%
%The Z-transform allows to compute the time series with the right sampling time in an efficient way.
%%%%%%%
%     }.
     \item 
     A Gaussian noise $W(t)$ is added to the time series, that is, a sequence of Gaussian random variables $W(i)$. The Gaussian noise has zero mean and variance computed as a fraction of the variance computed from the time series obtained in the previous step. In formulas:
     \begin{equation}
        S(i) = \mathcal{Z}^{-1} \left[ \mathcal{Z}\left[ S_{_{PT}} \right] \mathcal{Z} \left[ I \right] \, \right](i) + W(i) \ ; \quad  0 \le i\cdot T_s \le t_{max}=t_M
     \label{generated_signal}
     \end{equation}  
     %to the time series.
\end{enumerate}
The algorithm's functionality is illustrated in Figure \ref{fig:ExTimeSeriesGenerator}, 
%To evaluate the performance of the Rapid-Transition Finder (RTEF) in combination with the Event-Driven Diffusion Scaling (EDDiS) method, we chose to analyze time series with a single noise level. In particular, 
the noise variance is set to $10\%$ of the variance computed from the time series without noise.

\begin{figure}[H]
    \centering
    \includegraphics[scale=0.5]{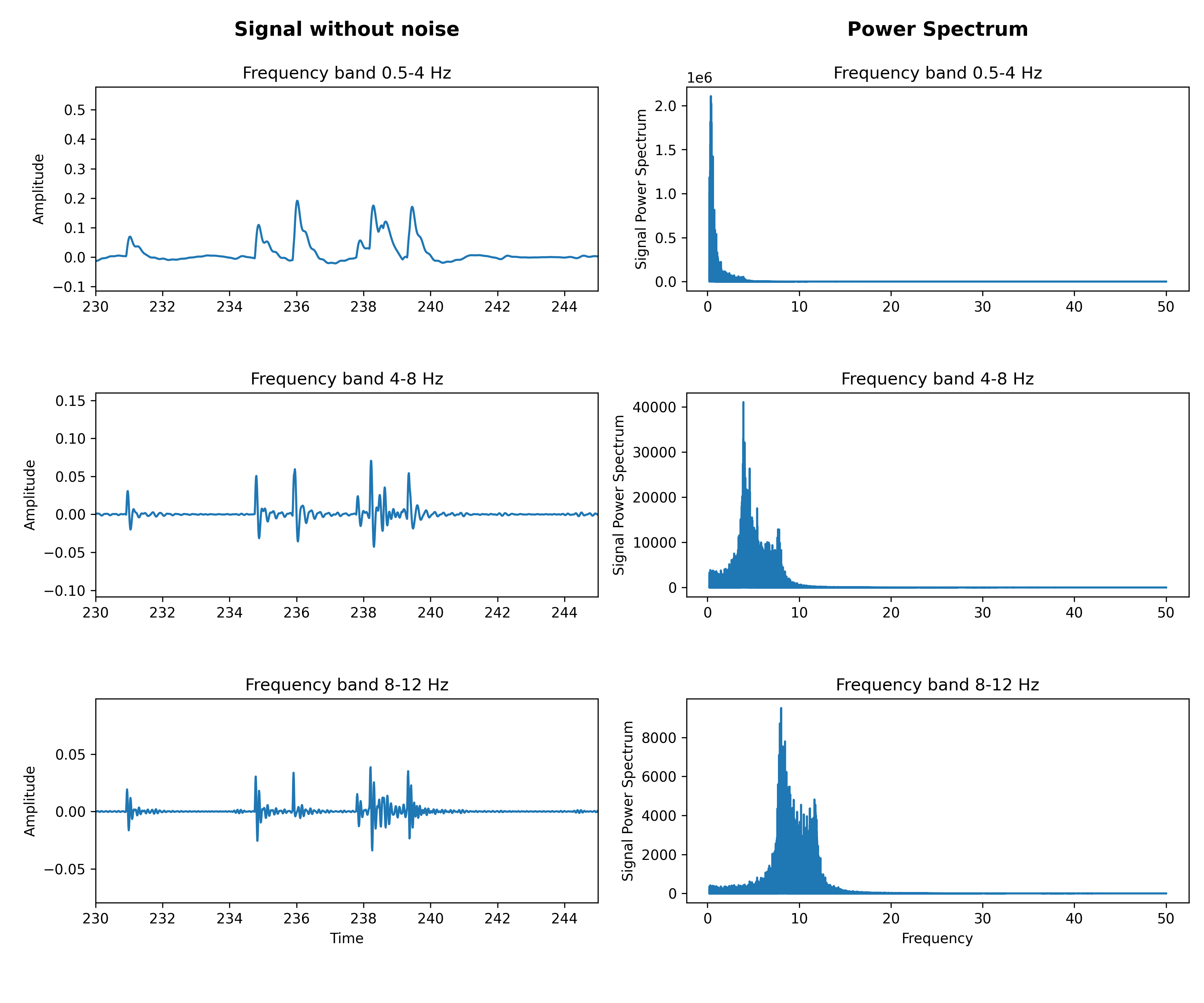}
    \caption{Example of a time series generated in different frequency bands. Left panels: generated time series; same event sequence as in the top right panel of Figure \ref{fig:ExTimeSeriesGenerator} (power-law, $\mu=2.5$). Right panels: power spectra corresponding to the time series reported in the left panels.}
    \label{fig:ExTimeSeriesBands}
\end{figure}

\noindent
We evaluate the performance of the RTEF by generating
%To our purposes, we generated 
time series in three distinct frequency bands $[\nu_{min},\nu_{max}]$: $[0.5,4]$Hz, $[4,8]$Hz, and $[8,12 ]$Hz. Figure \ref{fig:ExTimeSeriesBands} shows an example of the same time series generated in these different frequency bands starting from the same event sequence, without noise.
%%%%%%%%%%%%%%%%%%%%%%%%%%%%%%%%%%%%%%%%%%%%%%%%%%%%%%%%%%%%%%%%%%%%%%%%%%%%%
%%%%%%%%%%%%%%%%%%%%%% Plots for Time-Series Generator
%%%%%%%%%%%%%%%%%%%%%%%%%%%%%%%%%%%%%%%%%%%%%%%%%%%%%%%%%%%%%%%%%%%%%%%%%%%%%

%%%%%%%%%%%%%%%%%%%%%%%%%%%%%%%%%%%%%%%%%%%%%%%%%%%%%%%%%%%%%%%%%%%%%%%%%%%%%%%%%%%%%
%%%%%%%%%%%%%%%%%%%%%%%%%% Numerical Simulations and Results %%%%%%%%%%%%%%%%%%%%%%%%
%%%%%%%%%%%%%%%%%%%%%%%%%%%%%%%%%%%%%%%%%%%%%%%%%%%%%%%%%%%%%%%%%%%%%%%%%%%%%%%%%%%%%
\section{Numerical Simulations and Results}
\label{sec:num_sim}

% Set dei parametri utilizzati (per il segnale generato e per l'analisi)
\noindent
To test the RTEF algorithm we simulated different synthetic time series exploiting the model introduced in the previous Section \ref{sec:Signal_Generator}.
%This section discusses all the numerical simulations and introduces the most important results. 
% Start with talking about simulated signals
%We simulated different signals with the algorithm illustrated in Section \ref{sec:Signal_Generator}. 
%
We generated a total of $150$ signals for each IET-PDF, which are: power-law with $\mu = 2.3$, $\mu = 2.5$, $\mu = 2.7$; exponential with $r_p = 1$.
We recall that all IET sequences are constrained to a unitary mean IET. This is immediate in the exponential case, being $\langle \tau_p \rangle = 1/r_p = 1$. Conversely, in the power-law case, the parameter $T$ in Eqs. (\ref{powerlaw_wt}) and (\ref{PL_distribution}) must obey the relationship: $T/(\mu - 2) = \langle \tau_n \rangle = 1$.

\noindent
For each IET-PDF, $50$ different sample sets of IETs were drawn: $\{\tau_n;\ n=1,M\}$ ($M=20000$). The associated sequence of event time occurrence is then given by: $t_0=0$, $t_n=t_{n-1}+\tau_n,\ n\ge1$.
For each sample set of IET, we apply the points (4-5) of the algorithm for three different frequency bands of the signal: $[0.5,4]$Hz, $[4,8]$Hz, and $[8,12]$Hz. 
We finally add the Gaussian noise, as described in the point (6) of the previous section, to get the final temporal signal.

%events were drawn from the considered IET-PDF and, for each event sequence, the associated $50$ temporal signals were generated through the procedure introduced in previous Section \ref{sec:Signal_Generator}.
\noindent
We first evaluated the complexity indices $H$ and $\delta$ directly on
the IET samples, giving an estimation of relative errors due only to the random drawing, that is, to the random number generator, and to the best fit algorithm. This is obtained by applying the EDDiS, limited to the AJ walking rule, to the IET samples.
These estimated relative errors, which will be denoted in the following as {\it reference errors}, are presented in Table \ref{tab:WTsRefErrors}.
%presents the relative errors related to the random drawing of the events from each IET distribution.
This is a useful estimation of the statistical error associated with the random generator and the best fit procedure that can be used as a set of reference values for the combined RTEF-EDDiS pipeline.

\begin{figure}
    \centering
    \begin{subfigure}[b]{1\textwidth}
    \subcaption{85th Percentile}
    \centering  \includegraphics[width=.51\textwidth]{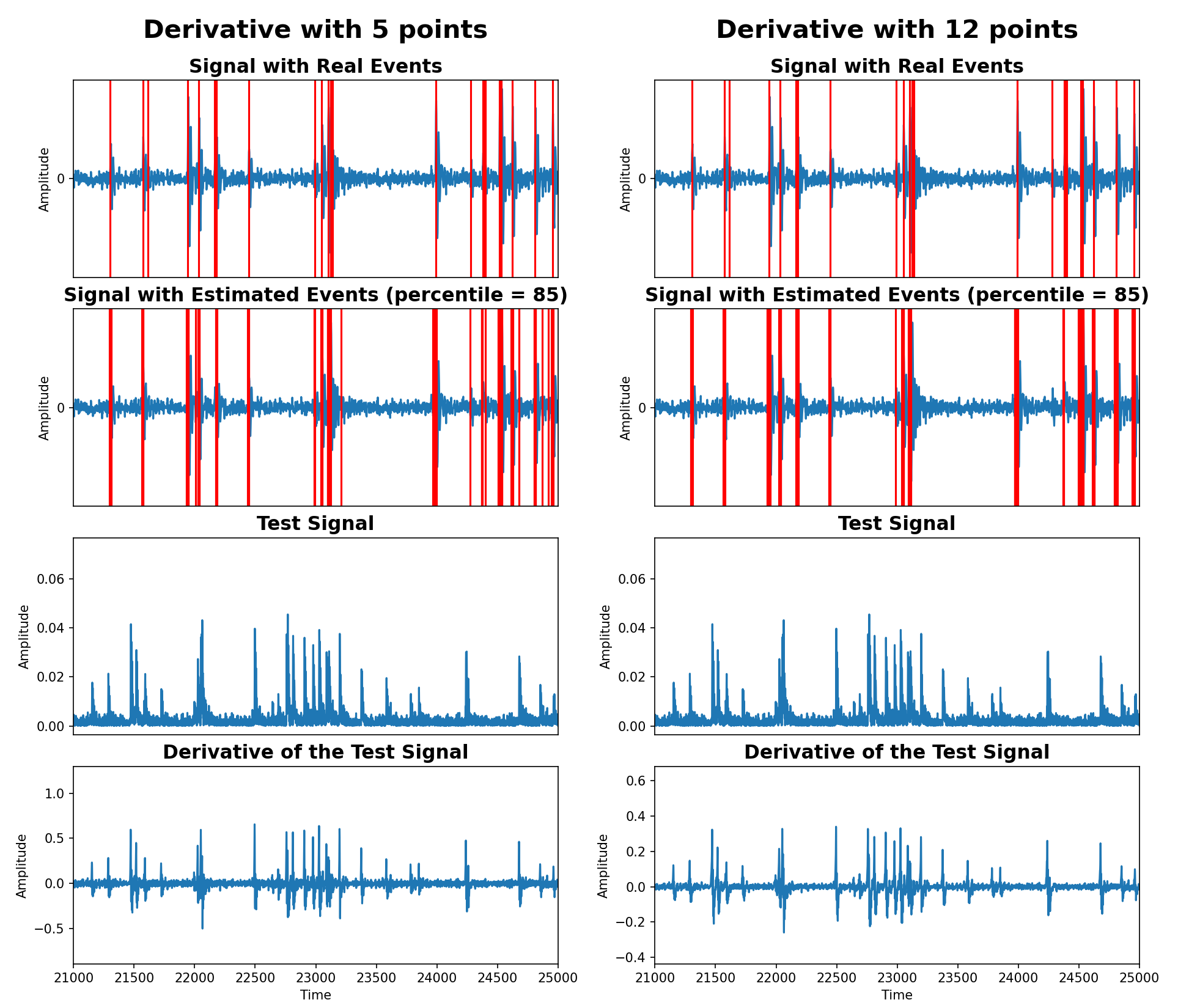}
    \end{subfigure}
    % \hspace{-.7cm}
    \begin{subfigure}[b]{1\textwidth}
    \subcaption{98th Percentile}
    \centering     
    \includegraphics[width=.51\textwidth]{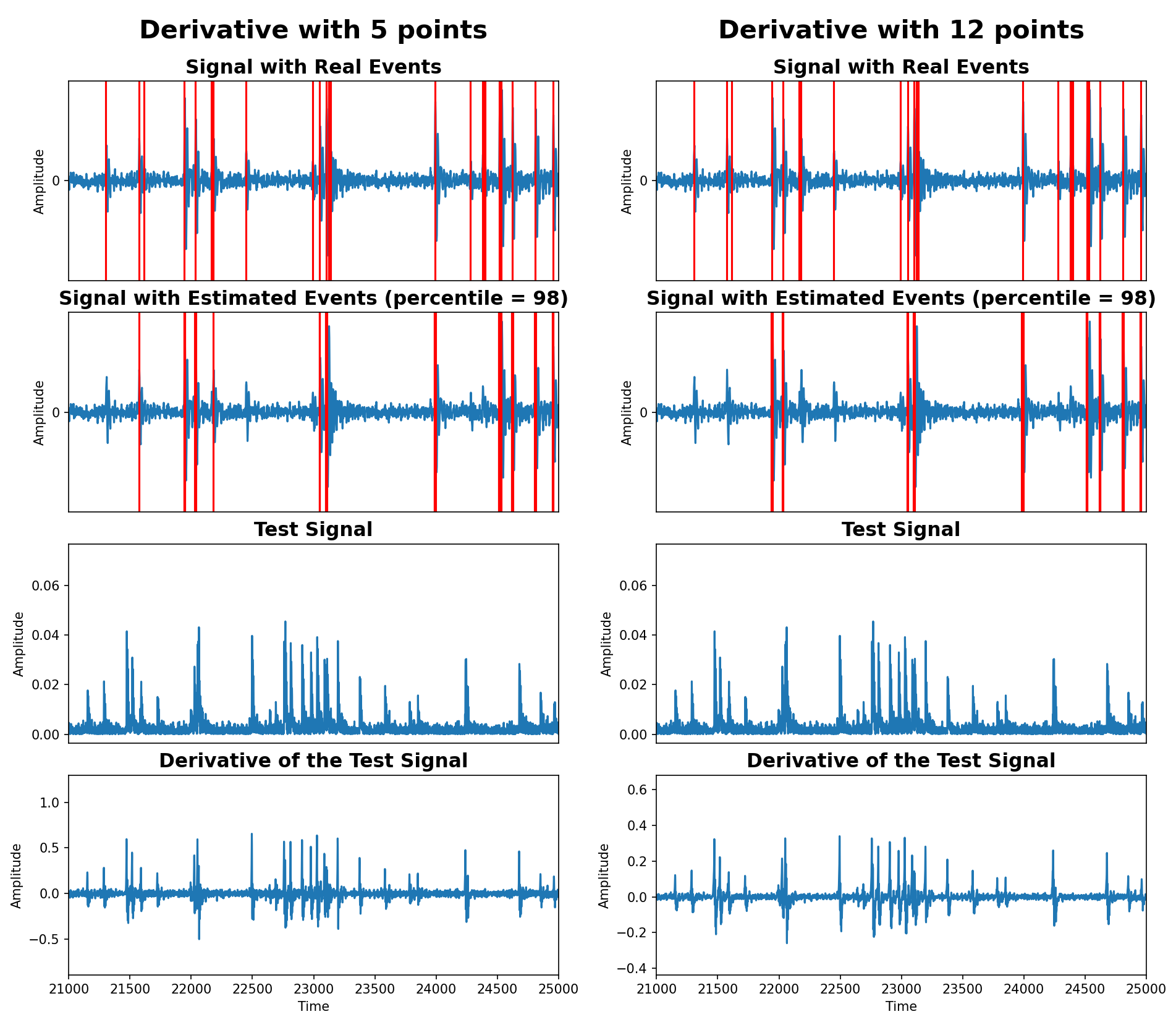}
    \end{subfigure}
    \caption{Example of application of the RTEF algorithm. The results are obtained using $5$ or $12$ symmetrical points to estimate the derivative for a signal generated in frequency band $[8,12]$Hz. On the first two top panels, vertical red lines indicate the real (above) or estimated (below) events. (a) RTEF results for the 85th percentile of the derivative distribution. (b) RTEF results for the 98th percentile of the derivative distribution.}
    \label{fig:RTEF_Alg_Ex}
\end{figure}

\begin{figure*}
    \centering
    \begin{subfigure}[b]{1\textwidth}
    \subcaption{85th Percentile}
    \centering
    \includegraphics[width=.51\textwidth]{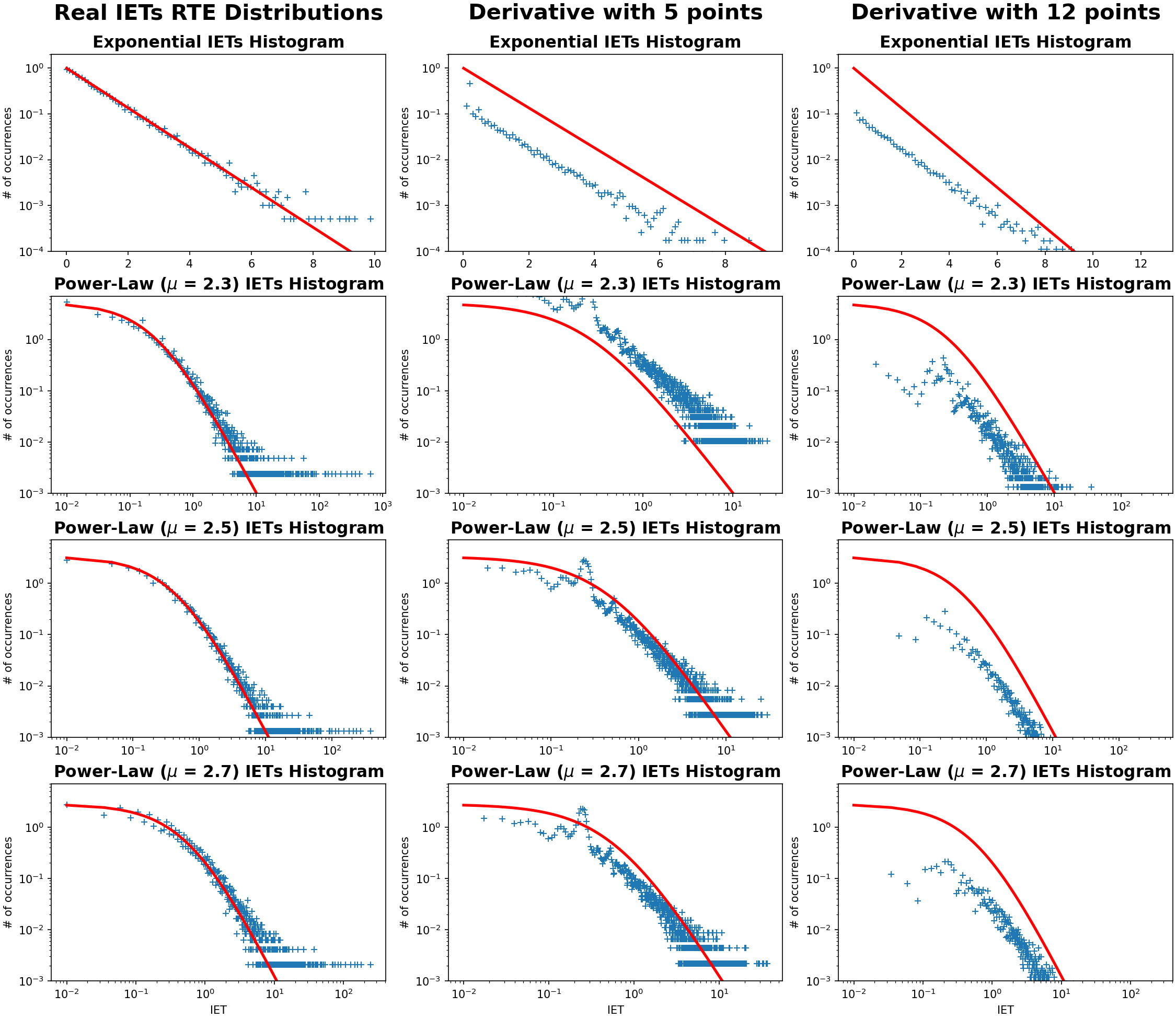}
 %    \vspace{-.3cm}
    \end{subfigure}
    \begin{subfigure}[b]{1\textwidth}
    \subcaption{98th Percentile}
    \centering
    \includegraphics[width=.51\textwidth]{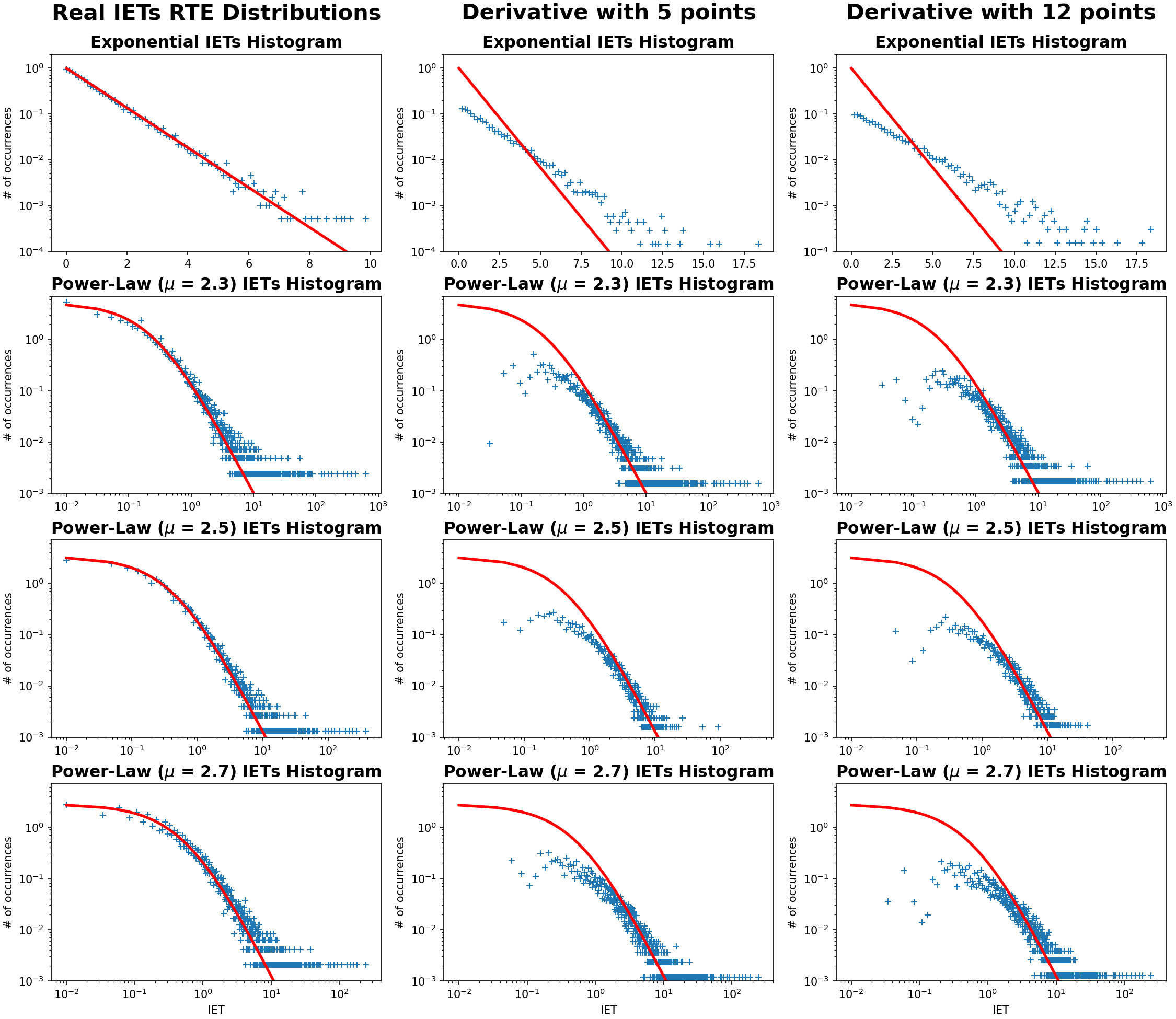}  
    \end{subfigure}
    \caption{Comparison of real IET distributions to the estimated ones for a signal generated in frequency band $[8, 12]$Hz and using $5$ or $12$ symmetrical points to estimate the derivative. The red lines represent the theoretical curves. (a) Comparison of real IETs to the estimated one from the 85th percentile of the derivative distribution. (b) Comparison of real IETs to the estimated one from the 98th percentile of the derivative distribution.}  \label{fig:WTs_Distributions_Found_vs_Real}
\end{figure*}

\vspace{.3cm}
\noindent
As illustrated before, we used two different time windows to estimate the derivative of the envelope of the absolute value of the signal. In Figure \ref{fig:RTEF_Alg_Ex} we display an example of how the RTEF algorithm works. In particular, these panels highlight that there are important differences in choosing $N_d = 5$ or $N_d=12$ symmetrical points to estimate the derivative. In fact, for all the percentile values that we used, the estimation of the derivative with $N_d=5$ gives more precise results than the one with $N_d=12$. Moreover, by passing from the $85$th to the $98$th percentiles, the number of false positives decreased, but some real events were missed. In Figure \ref{fig:RTEF_Alg_Ex} we report only the results with the $85$th and the $98$th percentiles, whereas other results with $90$th and $95$th percentiles are reported in the Supplementary Material. 
%
% Illustrare I risultati che si vedono ad occhio sulle distribuzioni dei WTs trovate rispetto a quelle reali
The IET distribution of the real and of the detected RTEs are similar but with some differences. In Figure \ref{fig:WTs_Distributions_Found_vs_Real} we can see that there are some differences in the IET-PDFs estimated by using $5$ or $12$ symmetrical points to estimate the derivative. The estimated distributions by using $5$ symmetrical points are more similar to the real one with respect to the one estimated with $12$ symmetrical points. By passing from the 85th percentile to the 98th percentile we noticed that the exponential distribution and the initial part of the power-law distributions got worse, while the tails of the power-law distributions became more similar to the real one.
Even if not trivial, this result is expected by observing that a higher percentile value select time events with greater signal steepness.
However, it is worth noting that the estimations of the
IET distributions are quite robust concerning the change in both $N_d$ and percentile.

%
% Parlare di come abbiamo stimato il diffusion scaling delle distribuzioni dei WTs (DFA, DE) e I risultati ottenuti riportando la tabella con gli errori medi e la deviazione standard di DFA/DE. 

\begin{figure}
    \centering
    \includegraphics[scale=0.4]{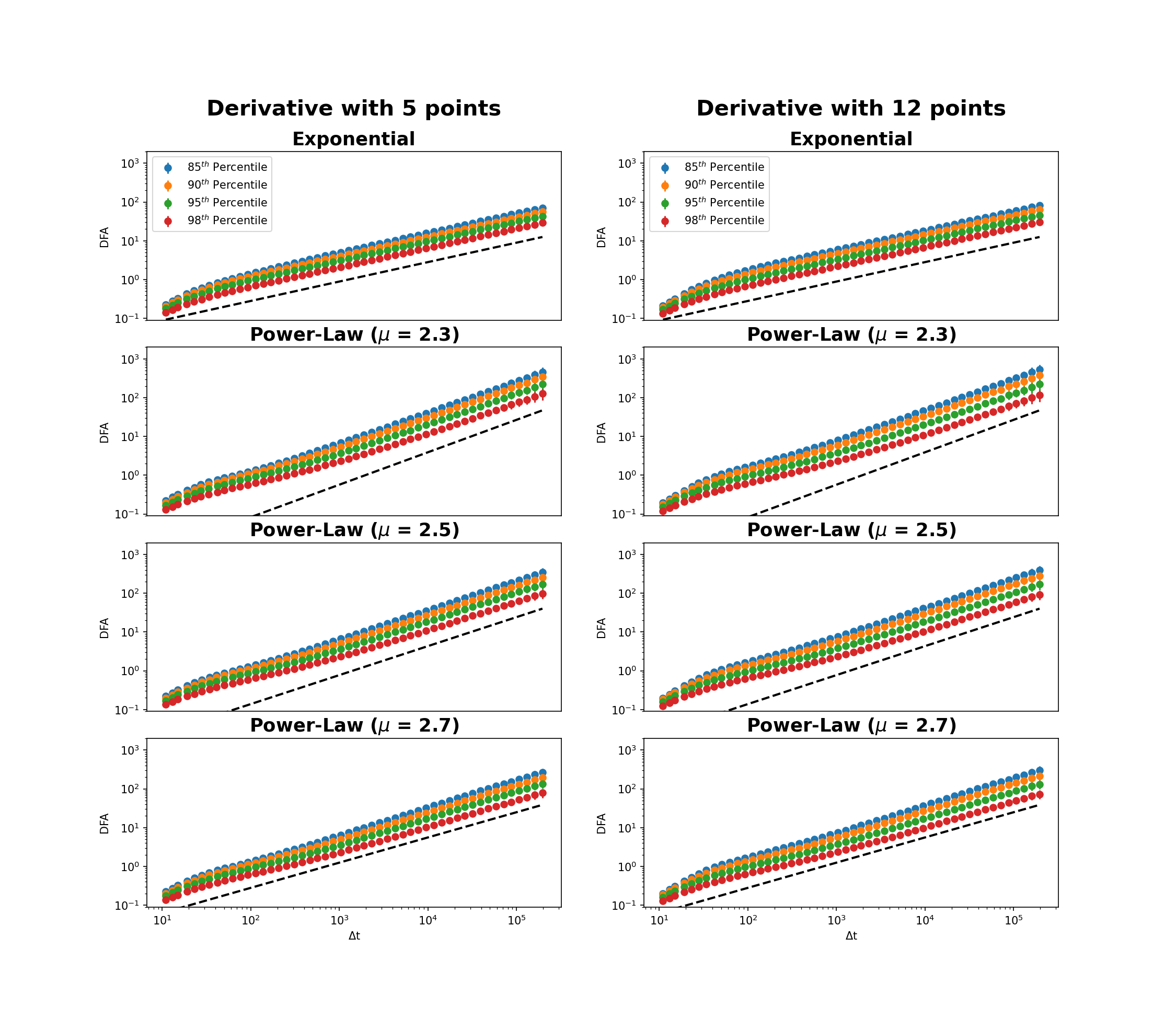}
    \caption{DFA mean curves with error bars for signals generated in frequency band $[8, 12]$Hz. Mean and error bars are computed over the $50$ IET samples. This figure compares the DFA curves obtained using $N_d=5$ and $N_d=12$ symmetrical points to estimate the derivative.
    %The blue dots are the results obtained by using the 85th percentile, the yellow dots are the results obtained by using the 90th percentile, the green dots are the results obtained by using the 95th percentile, red dots are the results obtained by using the 98th percentile, and 
    The dashed black lines represent the theoretical slope, that is: \(H=0.5\) for the Exponential Distribution, \(H=0.85\) for Power-Law distribution with $\mu = 2.3$, \(H= 0.75\) for $\mu = 2.5$, and \(H=0.65\) for $\mu = 2.7$.}
    \label{fig:DFAErrorbars}
\end{figure}

\begin{figure}
    \centering
    \includegraphics[scale=0.4]{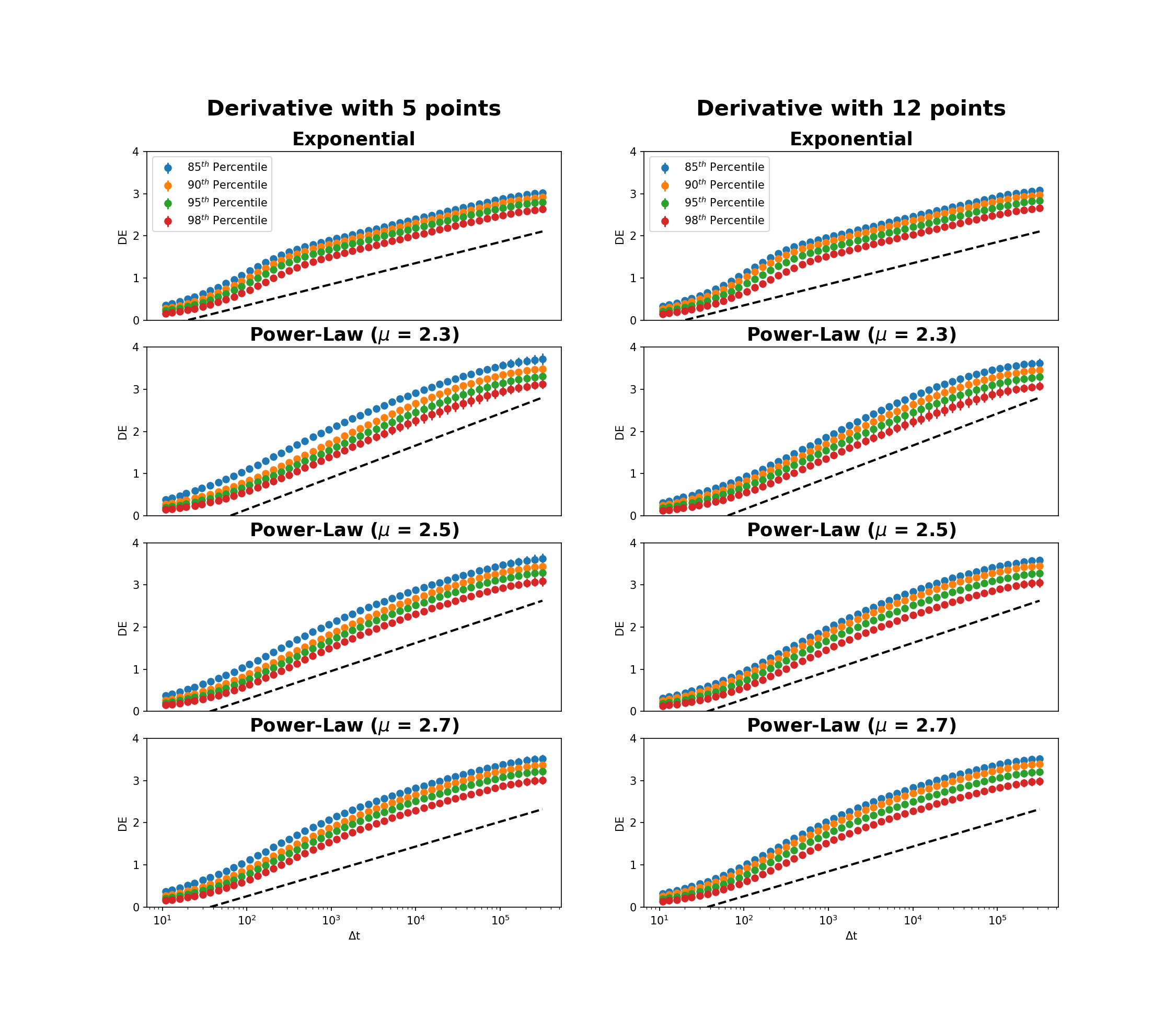}
    \caption{DE mean curves with error bars for signals generated in frequency band $[8, 12]$Hz. Mean and error bars are computed over the $50$ IET samples. This figure compares the DE curves obtained using $N_d=5$ and $N_d=12$ symmetrical points to estimate the derivative.
    %The blue dots are the results obtained by using the 85th percentile, the yellow dots are the results obtained by using the 90th percentile, the green dots are the results obtained by using the 95th percentile, red dots are the results obtained by using the 98th percentile, and 
    The dashed black lines represent the theoretical slope that is: \(\delta=0.5\) for the Exponential Distribution, \(\delta \simeq 0.77\) for Power-Law distribution with $\mu = 2.3$, \(\delta \simeq 0.67\) for $\mu = 2.5$, and \(\delta \sim 0.59\) for $\mu = 2.7$.}
    \label{fig:DEErrorbars}
\end{figure}

\noindent
To estimate the diffusion scaling parameter derived from the detected events, we employed the DFA and DE analyses. In Figure \ref{fig:DFAErrorbars} and Figure \ref{fig:DEErrorbars}, we report the average results of the DFA and DE analysis with error bars and with all the percentile values chosen. We report here the results obtained for signals generated in the frequency band $[8,12]$Hz, while the results obtained for signals generated in frequency bands $[0.5,4]$Hz and $[4,8]$Hz are reported in the Supplementary Materials. Tables \ref{tab:ExpDFADEErrors5pt}, \ref{tab:PL1DFADEErrors5pt}, \ref{tab:PL2DFADEErrors5pt}, and \ref{tab:PL3DFADEErrors5pt} present the Mean Relative Errors (MREs) with Standard Deviations (STDs) of estimated scaling parameters $H$ and $\delta$ obtained for each percentile and each frequency band, limited to $N_d=5$.
%symmetrical points for derivative estimation. 
We chose to report only the 5-point results due to their better performance compared to the 12-point results, which are included in the Supplementary Material.

%%%%%%%%%%%%%%%%%%%%%%%%%%%%%%%%%%%%%%%%%%%%%%%%%%%%%%%%%%%%%%%%%%%%%%%%%%%%%
%%%%%%%%%%%%%%%%%%%%%% Plots of Data Analysis
%%%%%%%%%%%%%%%%%%%%%%%%%%%%%%%%%%%%%%%%%%%%%%%%%%%%%%%%%%%%%%%%%%%%%%%%%%%%%
% RTEF Algorithm example

% WTs distributions real vs estimated

% DFA with Errorbars

% DE with Errorbars

%%%%%%%%%%%%%%%%%%%%%%%%%%%%%%%%%%%%%%%%%%%%%%%%%%%%%%%%
%%%% Tables DFA/DE Relative errors and Fit Relative Errors
%%%%%%%%%%%%%%%%%%%%%%%%%%%%%%%%%%%%%%%%%%%%%%%%%%%%%%%%
% Real WTs Distributions Relative Error Tables
\begin{table}[H]
\centering   
    \begin{subtable}[t]{1\linewidth}
    \centering
    \begin{tabular}[t]{c|c|c} 
    \textbf{Case} & \textbf{MRE H ($\pm$ STD)} & \textbf{MRE $\delta$ ($\pm$ STD)}\\
      \hline
      \hline
      Exponential & 0.0052 ($\pm$ 0.0042) & 0.024 ($\pm$ 0.011) \\
      \relax
      Power-Law ($\mu = 2.3$) & 0.088 ($\pm$ 0.040) & 0.040 ($\pm$ 0.035) \\
      \relax
      Power-Law ($\mu = 2.5$) & 0.040 ($\pm$ 0.030) & 0.062 ($\pm$ 0.021) \\
      \relax
      Power-Law ($\mu = 2.7$) & 0.070 ($\pm$ 0.040) & 0.15 ($\pm$ 0.030) \\
    \end{tabular}
    \caption{DFA and DE Mean Relative Errors}
    \label{tab:DFADERefErrors}
    \end{subtable}
    \vspace{15pt}
    \begin{subtable}[t]{1\linewidth}
    \centering
    \begin{tabular}[t]{c|c|c} 
    \textbf{Case} & \textbf{MRE $\mu$ or $r_p$ ($\pm$ STD)}& \textbf{Min and Max Relative Error}\\
      \hline
      \hline
      Exponential & 0.019 ($\pm$ 0.012) & [0.001, 0.044] \\
      \relax
      Power-Law ($\mu = 2.3$) & 0.068 ($\pm$ 0.055) & [0.0031, 0.25] \\
      \relax
      Power-Law ($\mu = 2.5$) & 0.061 ($\pm$ 0.046) & [0.0002, 0.22] \\
      \relax
      Power-Law ($\mu = 2.7$) & 0.051 ($\pm$ 0.041) & [0.0015, 0.16] \\
    \end{tabular}
    \caption{Fit Mean Relative Errors}
    \label{tab:RefErrorsFits}
    \end{subtable}
\vspace{-.7cm}
    \caption{{\it Reference errors} for complexity analysis
    % DFA and DE 
    carried out directly on the IET samples (average and standard deviation are computed on the $50$ IET samples). (a) Mean Relative Errors (MREs) with Standard Deviation (STD) of DFA and DE; (b) MREs and STDs on $\mu$ for Power-Law and $r_p$ for exponential distributions computed on the $50$ IET samples.}
    \label{tab:WTsRefErrors}
\end{table}

% Exp Case detected Events
\newpage
\begin{table}[H]
    \centering   
    \begin{tabular}[t]{c|c|c|c} 
    \textbf{Frequency Band} & \textbf{Percentile} & \textbf{MRE H ($\pm$ STD)} & \textbf{MRE $\delta$ ($\pm$ STD)}\\
      \hline
      \hline
      \multirow{4}{*}{[0.5, 4]} & 85 & 0.033 ($\pm$ 0.016) & 0.056 ($\pm$ 0.050) \\
      & 90 & 0.038 ($\pm$ 0.017) & 0.061 ($\pm$ 0.051) \\
      & 95 & 0.037 ($\pm$ 0.016) & 0.081 ($\pm$ 0.065) \\
      & 98 & 0.033 ($\pm$ 0.018) & 0.12 ($\pm$ 0.084) \\
      \hline
      \relax
      \multirow{4}{*}{[4, 8]} & 85 & 0.031 ($\pm$ 0.015) & 0.073 ($\pm$ 0.052) \\
      & 90 & 0.014 ($\pm$ 0.010) & 0.082 ($\pm$ 0.054) \\
      & 95 & 0.010 ($\pm$ 0.010) & 0.085 ($\pm$ 0.064) \\
      & 98 & 0.012 ($\pm$ 0.010) & 0.11 ($\pm$ 0.083) \\
      \hline
      \relax
      \multirow{4}{*}{[8, 12]}  & 85 & 0.028 ($\pm$ 0.015) & 0.069 ($\pm$ 0.052) \\
      & 90 & 0.013 ($\pm$ 0.010) & 0.072 ($\pm$ 0.048) \\
      & 95 & 0.010 ($\pm$ 0.010) & 0.080 ($\pm$ 0.054) \\
      & 98 & 0.012 ($\pm$ 0.010) & 0.091 ($\pm$ 0.069) \\
    \end{tabular}
    \caption{MREs and STDs of DFA and DE computed from detected events in signals driven by exponential IET distributions (MREs and STDs computed from the $50$ IET samples). All parameter configurations of the RTEF algorithm are reported for comparison. Results were obtained by using $N_d=5$.}
    \label{tab:ExpDFADEErrors5pt}
\end{table}

% PL Case (mu = 2.3) Detected Events
\begin{table}[H]
    \centering   
    \begin{tabular}[t]{c|c|c|c} 
    \textbf{Frequency Band} & \textbf{Percentile} & \textbf{MRE H ($\pm$ STD)} & \textbf{MRE $\delta$ ($\pm$ STD)}\\
      \hline
      \hline
      \multirow{4}{*}{[0.5, 4]} & 85 & 0.073 ($\pm$ 0.044) & 0.060 ($\pm$ 0.011) \\
      & 90 & 0.079 ($\pm$ 0.049) & 0.10 ($\pm$ 0.013) \\
      & 95 & 0.10 ($\pm$ 0.064) & 0.12 ($\pm$ 0.014) \\
      & 98 & 0.10 ($\pm$ 0.052) & 0.044 ($\pm$ 0.017) \\
      \hline
      \relax
      \multirow{4}{*}{[4, 8]} & 85 & 0.065 ($\pm$ 0.044) & 0.17 ($\pm$ 0.040) \\
      & 90 & 0.077 ($\pm$ 0.050) & 0.13 ($\pm$ 0.033) \\
      & 95 & 0.081 ($\pm$ 0.046) & 0.090 ($\pm$ 0.030) \\
      & 98 & 0.10 ($\pm$ 0.056) & 0.040 ($\pm$ 0.024) \\
      \hline
      \relax
      \multirow{4}{*}{[8, 12]}  & 85 & 0.064 ($\pm$ 0.040) & 0.17 ($\pm$ 0.036) \\
      & 90 & 0.073 ($\pm$ 0.043) & 0.12 ($\pm$ 0.034) \\
      & 95 & 0.077 ($\pm$ 0.043) & 0.090 ($\pm$ 0.031) \\
      & 98 & 0.092 ($\pm$ 0.050) & 0.046 ($\pm$ 0.025) \\
    \end{tabular}       
    \caption{MREs and STDs of DFA and DE computed from detected events in signals driven by Power-Law IET distribution with $\mu = 2.3$ and $T = 0.3$ (MREs and STDs computed from the $50$ IET samples). All parameter configurations of the RTEF algorithm are reported for comparison. Results were obtained by using $N_d=5$.}
    \label{tab:PL1DFADEErrors5pt}    
\end{table}

% PL Case (mu = 2.5) Detected Events
\begin{table}[H]
    \centering   
    \begin{tabular}[t]{c|c|c|c} 
    \textbf{Frequency Band} & \textbf{Percentile} & \textbf{MRE H ($\pm$ STD)} & \textbf{MRE $\delta$ ($\pm$ STD)}\\
      \hline
      \hline
      \multirow{4}{*}{[0.5, 4]} & 85 & 0.073 ($\pm$ 0.035) & 0.17 ($\pm$ 0.055) \\
      & 90 & 0.080 ($\pm$ 0.031) & 0.19 ($\pm$ 0.076) \\
      & 95 & 0.095 ($\pm$ 0.036) & 0.23 ($\pm$ 0.074) \\
      & 98 & 0.12 ($\pm$ 0.054) & 0.19 ($\pm$ 0.014) \\
      \hline
      \relax
      \multirow{4}{*}{[4, 8]} & 85 & 0.041 ($\pm$ 0.014) & 0.24 ($\pm$ 0.092) \\
      & 90 & 0.061 ($\pm$ 0.030) & 0.28 ($\pm$ 0.065) \\
      & 95 & 0.063 ($\pm$ 0.042) & 0.24 ($\pm$ 0.044) \\
      & 98 & 0.071 ($\pm$ 0.045) & 0.17 ($\pm$ 0.018) \\
      \hline
      \relax
      \multirow{4}{*}{[8, 12]}  & 85 & 0.040 ($\pm$ 0.020) & 0.24 ($\pm$ 0.092) \\
      & 90 & 0.052 ($\pm$ 0.033) & 0.28 ($\pm$ 0.060) \\
      & 95 & 0.051 ($\pm$ 0.033) & 0.24 ($\pm$ 0.043) \\
      & 98 & 0.062 ($\pm$ 0.042) & 0.18 ($\pm$ 0.020) \\
    \end{tabular} 
    \caption{MREs and STDs of DFA and DE computed from detected events in signals driven by Power-Law IET distribution with $\mu = 2.5$ and $T = 0.5$ (MREs and STDs computed from the $50$ IET samples). All parameter configurations of the RTEF algorithm are reported for comparison. Results were obtained by using $N_d=5$.}
    \label{tab:PL2DFADEErrors5pt}  
\end{table}

% PL Case (mu = 2.7) Detected Events
\begin{table}[H]
    \centering   
    \begin{tabular}[t]{c|c|c|c} 
    \textbf{Frequency Band} & \textbf{Percentile} & \textbf{MRE H ($\pm$ STD)} & \textbf{MRE $\delta$ ($\pm$ STD)}\\
      \hline
      \hline
      \multirow{4}{*}{[0.5, 4]} & 85 & 0.020 ($\pm$ 0.010) & 0.14 ($\pm$ 0.098) \\
      & 90 & 0.023 ($\pm$ 0.014) & 0.13 ($\pm$ 0.090) \\
      & 95 & 0.018 ($\pm$ 0.013) & 0.18 ($\pm$ 0.10) \\
      & 98 & 0.051 ($\pm$ 0.018) & 0.25 ($\pm$ 0.088) \\
      \hline
      \relax
      \multirow{4}{*}{[4, 8]} & 85 & 0.071 ($\pm$ 0.021) & 0.16 ($\pm$ 0.090) \\
      & 90 & 0.029 ($\pm$ 0.015) & 0.23 ($\pm$ 0.11) \\
      & 95 & 0.028 ($\pm$ 0.011) & 0.23 ($\pm$ 0.11) \\
      & 98 & 0.055 ($\pm$ 0.030) & 0.24 ($\pm$ 0.087) \\
      \hline
      \relax
      \multirow{4}{*}{[8, 12]}  & 85 & 0.066 ($\pm$ 0.015) & 0.17 ($\pm$ 0.095) \\
      & 90 & 0.024 ($\pm$ 0.012) & 0.23 ($\pm$ 0.011) \\
      & 95 & 0.030 ($\pm$ 0.010) & 0.23 ($\pm$ 0.011) \\
      & 98 & 0.049 ($\pm$ 0.027) & 0.23 ($\pm$ 0.095) \\
    \end{tabular}
    \caption{MREs and STDs of DFA and DE computed from detected events in signals driven by Power-Law IET distribution with $\mu = 2.7$ and $T = 0.7$ (MREs and STDs computed from the $50$ IET samples). All parameter configurations of the RTEF algorithm are reported for comparison. Results were obtained by using $N_d=5$.}
    \label{tab:PL3DFADEErrors5pt}
\end{table}

%%%%%%%%%%%%%%%%%%%%%%%%%%%%%%%%%%%%%%%%%%%%%%%
%%%%%%%%%%%%%%%%%%%%%%%%%%%%%%%%%%%%%%%%%%%%%%%
%%%%%%%%%% Discussion 
%%%%%%%%%%%%%%%%%%%%%%%%%%%%%%%%%%%%%%%%%%%%%%%
%\subsection{Discussion of numerical results}
\section{Discussion of results}
\label{sec:discussion}

% Start by discussing in general the importance of the derivative estimation parameter choice to the algorithm performances

\noindent
We here give a detailed discussion of the numerical results regarding the application of EDDiS algorithm to the 
IET samples and of the RTEF-EDDiS pipeline to the artificial signals driven by the same IET samples.

\subsection{EDDiS analysis of IET samples and reference errors}

\noindent
Reference errors are due to:
finite size of the statistical sample; limitations of the random number generator; and intrinsic errors of the best-fit procedure. These errors were 
estimated by applying the EDDiS algorithm to the IET samples. The ideal values of $H$, $\delta$ and $\mu$ are
given by Eqs. (\ref{H_AJ}) and (\ref{delta_AJ}) and are compared to the corresponding estimated values by applying the RE formula, Eq. (\ref{relative_error}), and taking the average over the $50$ samples to get the MRE.

\noindent
Table \ref{tab:DFADERefErrors} shows the mean relative errors in the estimation of the scaling factors \(H\) and \(\delta\) obtained by applying DFA and DE algorithms on the real events drawn from all considered IET-PDFs. Notably, the relative errors for both scaling factors across all IET distributions remain below \(10\%\), except for the \(\delta\) scaling factor estimated for the inverse power-law distribution with \(\mu = 2.7\), which reaches \(15\%\). 
\noindent
In particular, the estimated $H$ values consistently exhibited a relative error below $8.8\%$, with the lowest error occurring for the exponential distribution and the highest for the power-law distribution with $\mu = 2.3$. 
Conversely, the $\delta$ estimated values %generally 
have higher MRE than $H$ values, reaching a maximum of $15\%$ for the power-law distribution with $\mu = 2.7$. There is only one exception: the power-law distribution with $\mu = 2.3$, where the relative error of $\delta$ was lower than that of $H$ ($4.0\%$ compared to $8.8\%$). 

\noindent
%Conversely, 
Table \ref{tab:RefErrorsFits} highlights that the mean relative errors in the direct $\mu$ (or $r_p$) evaluation,
i.e., from the fit of IET-PDFs, for all the cases do not pass the \(7\%\) on average. 
%The accuracy in $\mu$ evaluation becomes worst as $\mu$ decreases, while the best accuracy is again obtained for the evaluation of $r_p$ in the exponential case.
%
As expected, the best performance is given by the evaluation of $r_p$ in the exponential case, while the 
accuracy in $\mu$ evaluation increases (MRE decreases) as $\mu$ increases, thus giving the worst performance for the slower power-law decay, i.e., $\mu=2.3$. 
%The unique exception is given by the estimated $\delta$ for the case $\mu=2.7$, whose error can be attributed to the limitation of an automatic power-law fit procedure. 
%
%Examining the relative errors on the $\mu$ (or $r_p$) fits, we found that they consistently remained below $7\%$ on average. 
Moreover, an analysis of the confidence intervals revealed a maximum relative error of $25\%$ in the worst-case scenario and of $0.44\%$ in the best-case scenario. In contrast, the minimum relative errors were assessed at $0.31\%$ in the worst-case scenario and $0.02\%$ in the best-case scenario.

\noindent
In general, the $\delta$ and $H$ evaluations have the highest accuracy for the exponential case. For the power-law case, $\delta$ is better estimated for smaller $\mu$, while the accuracy of $H$ estimation is not monotonic with $\mu$, giving the best result for $\mu=2.5$ ($4\%$).

%%%%%%%%%%%%%%%%%%%%%%%%%%%%%%%%%%%
\subsection{RTEF-EDDiS analysis of event-driven artificial signals}

\noindent
The artificial signals generated according to the model of Section \ref{sec:Signal_Generator} were processed with the RTEF algorithm and the detected events were exploited to estimate the scaling exponents $H$ and $\delta$ through the EDDiS algorithm.
We found that the RTEF algorithm's performance is primarily affected by the number $N_d$ of symmetrical time points used for derivative estimation of the absolute value of the signal's envelope. In particular, we highlight that the case $N_d=5$ generally outperforms the case $N_d=12$.
%for the RTEF algorithm's performance.
For this reason, we here focus the discussion on the case $N_d=5$, while interested readers can find the results obtained using $N_d=12$ in the supplementary material.

% Illustrate that better results are obtained for percentile values that have a large number of false positives (talk about the figure counts vs. percentiles)
\noindent
Another crucial factor affecting the RTEF-EDDiS pipeline's performance is the percentile value used to select the  RTEs.
%truncate the derivative distribution, which determines the RTEs. 
As expected, the lower the percentile value, the more false positives are found. 
%
% Counting Detected Events vs Percentiles
\begin{figure}[]
    \centering
    \includegraphics[scale=0.4]{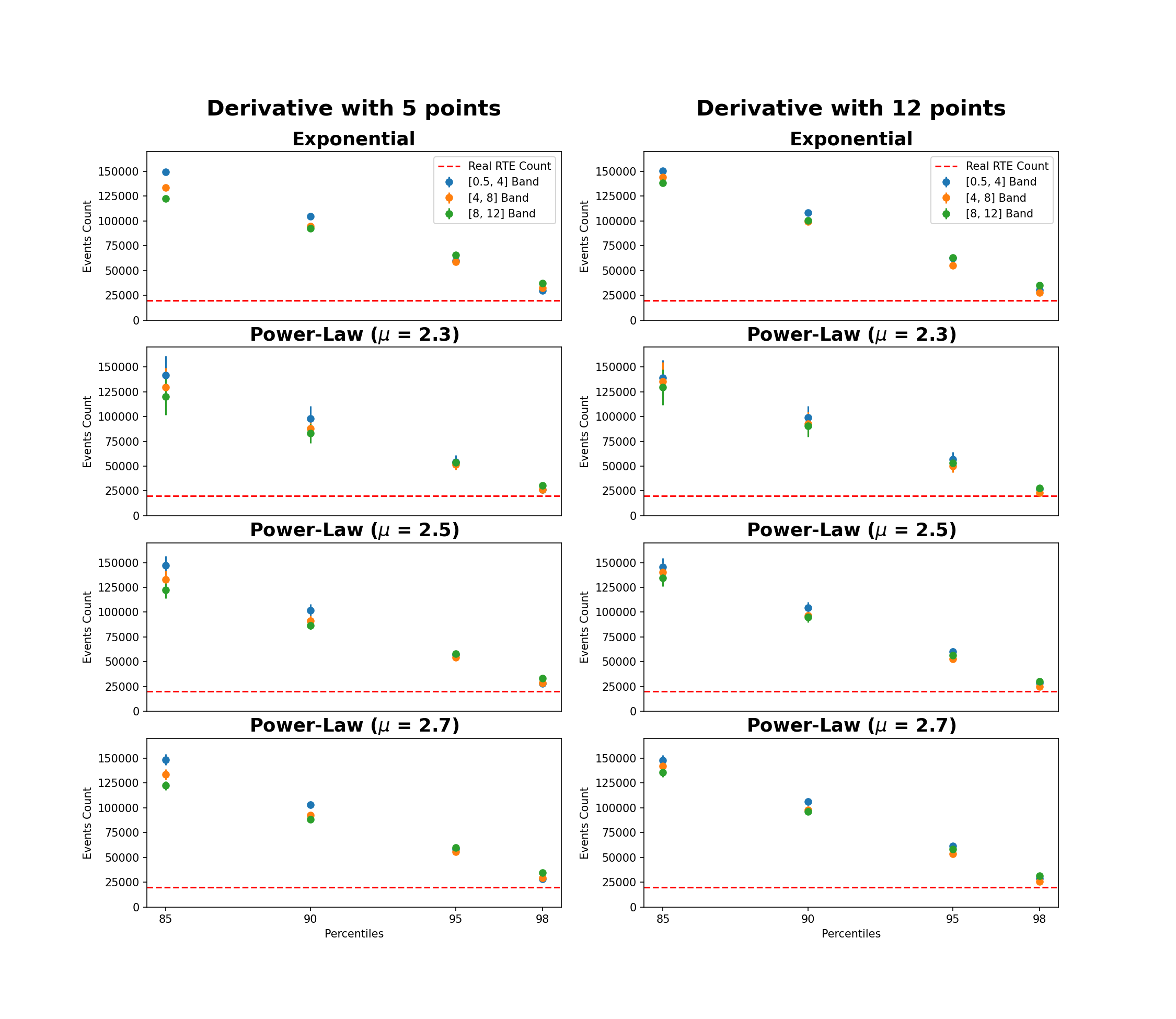}
    \caption{Mean estimated RTE counts with error bars for each IET distribution and frequency band vs percentile values. Left panels and right panels refer to $N_d=5$ and $N_d=12$, respectively.
    %The blue dots are the events count for the signal in the frequency band $[0.5, 4Hz]$, the yellow dots are the events count for the signal in the frequency band $[4, 8Hz]$, the green dots are the events count for signal in the frequency band $[8, 12Hz]$, and
    The dashed red lines represent the number of real RTEs that is equal to $20000$.}
    \label{fig:CountvsPerc}
\end{figure}
\noindent
From Figure \ref{fig:CountvsPerc} we can see that the number of detected RTEs spanned from 
%values with overestimation errors of the order of   $20-25\%$
two times 
to almost ten times the number of real RTEs going
from $95th$ to $85th$ percentile values, thus revealing a very large number of false positives. Only the $98$th percentile gives
RTE counts comparable with the real number of RTEs ($20000$), but still with an overestimation error of about $20-30\%$.

%%%%%%%%%%%%%%%%%%%%%%%%%%%%%%%%%%%
\subsubsection{Exponential case}

\noindent
%  Table exp
For the exponential case,
%, giving Poisson distributed events, 
summarized in Table 
\ref{tab:ExpDFADEErrors5pt}, 
%an opposite trend is observed, that is, 
the accuracy in $H$ improves as the percentile increases, or remains almost constant for the band $[0.5,4]$. This last band is also that giving the worst MREs, which however remain below $4\%$. 
%This behavior in the accuracy seems to be compatible with the intuition of a better accuracy of event detection when the percentile increases.
At variance with $H$, in all bands, the accuracy in the $\delta$ evaluation decreases as the percentile increases.
%While relative errors for $\delta$ are slightly lower for signals within the frequency band $[0.5,4]$Hz, they are still comparable to those for signals in other frequency bands. 

%%%%%%%%%%%%%%%%%%%%%%%%%%%%%%%%%%%
\subsubsection{Power-law $\mu=2.3$}

\noindent
For the power-law distribution with $\mu = 2.3$, we find the counter-intuitive result that the relative errors of $H$ 
%after the RTE detection 
are generally smaller than the reference errors of Table \ref{tab:WTsRefErrors}.
Further, accuracy improves as percentile decreases, a condition corresponding to a much larger number of false positives in the RTE detection.
%Specifically, percentiles less than $95$ often yield better results across most frequency bands. 
Additionally, the variation in $H$ relative errors between frequency bands is minimal. 
Conversely, the relative errors of $\delta$ 
%after the RTEs extraction 
are typically higher than reference errors and generally decrease as the percentile increases. In the band $[0.5,4]$ a non-monotonic behavior is seen, but the best accuracy is still found at the $98$th percentile with a $4.4\%$ relative error.
%better results generally obtained using percentile values greater than $90$.
Errors in $\delta$ are slightly larger than those in $H$.

%%%%%%%%%%%%%%%%%%%%%%%%%%%%%%%%%%%
\subsubsection{Power-law $\mu=2.5$}

\noindent
For the power-law distribution with $\mu=2.5$, we find that the relative errors of $H$ and $\delta$ are generally higher than the reference errors. Similarly to previous cases, the differences in the relative errors between frequency bands are minimal, even if we found slightly better results for $H$ in the $[4,8]$Hz and $[8,12]$Hz bands.

\noindent
Regarding $\delta$, the accuracy is non-monotonic with the percentile. The best results are given by the $98$th percentile in the $[4,8]$ and $[8,12]$ bands, and by the $85$th percentile in the $[0.5,4]$ band.
Interestingly, the errors in $\delta$ are much higher than those in $H$.
%Additionally, we note that the worst results for $H$ estimation occurred at the $98$th percentile, whereas $\delta$ results are similar across all percentiles.

%%%%%%%%%%%%%%%%%%%%%%%%%%%%%%%%%%%
\subsubsection{Power-law $\mu=2.7$}

\noindent
Surprisingly, for the power-law distribution with $\mu = 2.7$, the relative errors of $H$ are consistently lower than reference errors given in Table \ref{tab:DFADERefErrors}. Conversely, the $\delta$ relative errors are generally greater or equal to the corresponding reference errors. Regarding frequency band differences, minimal variations were observed in the relative errors of both $H$ and $\delta$ between $[4,8]$ and $[8,12]$ bands, while the $[0.5,4]$Hz band exhibits slightly better results. 
Finally, the $90$th and $95$th percentiles yielded the best results for $H$ estimation, while the 
$\delta$ estimation improves as the percentile decreases, reaching the best performance at the
$85$th percentile. Similarly to the case $\mu=2.5$, the errors in $\delta$ are much higher than those in $H$.
%

%%%%%%%%%%%%%%%%%%%%%%%%%%%%%%%%%%%
%%%%%%%%% Conclusions %%%%%%%%%%%%%%%%%%%%%%%%%%%%%%%%%%%
\section{Concluding remarks}
\label{sec:conclusions}

\noindent
In this work, we evaluated the effect of an event detection algorithm on the estimation accuracy of event-based complexity indices $\mu$, $H$ and $\delta$.
In the present stud,y we were not interested in the efficiency of single-event detection, but in
the possibility of getting a good estimation of the genuine temporal complexity.
In particular, our main interest lies in the evaluation of signals driven by complex events, that is, events with inverse power-law IET distributions.

\begin{figure}[H]
    \centering
    \includegraphics[scale=0.4]{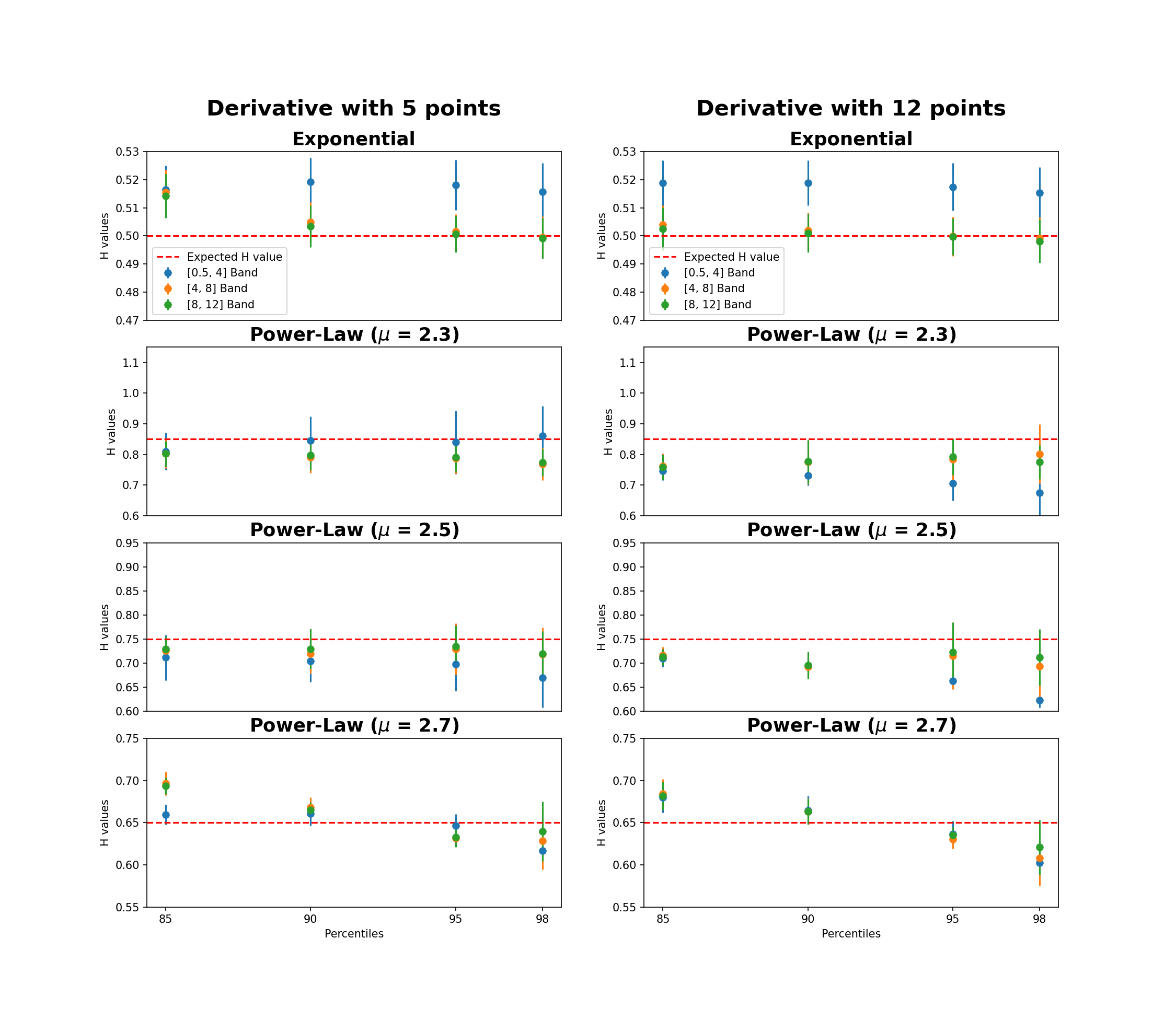}
    \caption{Mean estimated $H$ values with error bars for each IET distribution and frequency band vs percentile values. 
    %This figure compares the average $H$ values estimated with the respective error bars to the percentile value used to find them. 
    Both $N_d = 5$ and $N_d = 12$ cases are reported. 
    %Moreover, we want to show the difference between the H values estimated using 5 or 12 symmetrical points to estimate the derivative. The blue dots are the H values for signal in the frequency band $[0.5, 4Hz]$, the yellow dots are the H values for signal in the frequency band $[4, 8Hz]$, the green dots are the H values for signal in the frequency band $[8, 12Hz]$, and 
    The dashed red lines represent the theoretical $H$ values: \(0.5\) for the Exponential Distribution, \(0.85\) for Power-Law distribution with $\mu = 2.3$, \(0.75\) for $\mu = 2.5$, and \(0.65\) for $\mu = 2.7$.}
    \label{fig:HvsPerc}
\end{figure}

\begin{figure}[H]
    \centering
    \includegraphics[scale=0.4]{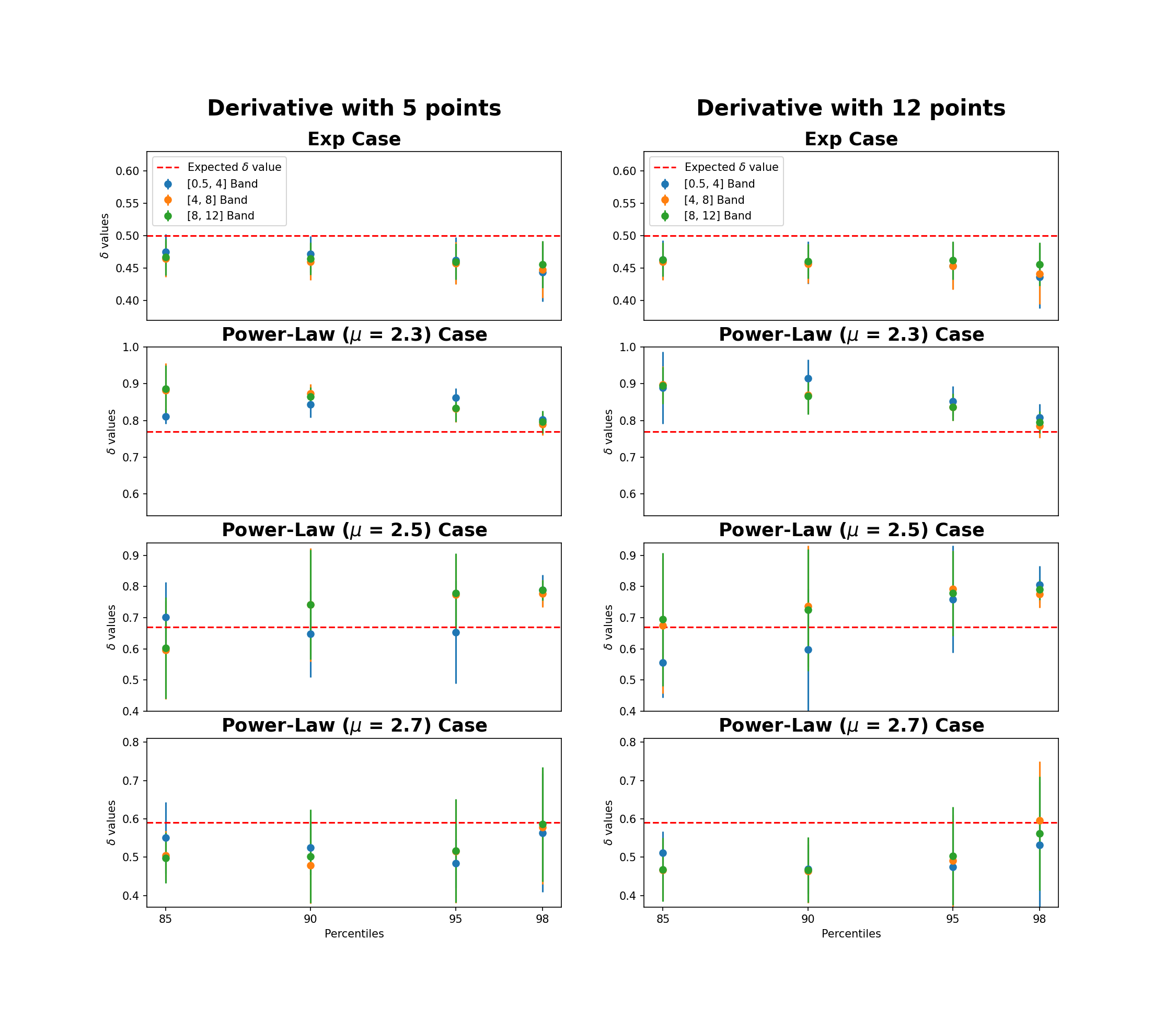}
    \caption{Mean estimated $\delta$ values with error bars for each IET distribution and frequency band vs percentile values. %This figure compares the average $\delta$ values estimated with the respective error bars to the percentile value used to find them. 
    Both $N_d = 5$ and $N_d = 12$ cases are reported.
    %Moreover, we want to show the difference between the $\delta$ values estimated using 5 or 12 symmetrical points to estimate the derivative. The blue dots are the $\delta$ values for the signal in the frequency band $[0.5, 4Hz]$, the yellow dots are the $\delta$ values for the signal in the frequency band $[4, 8Hz]$, the green dots are the $\delta$ values for signal in the frequency band $[8, 12Hz]$, and 
    The dashed red lines represent the theoretical $\delta$ values: \(0.5\) for the Exponential Distribution, \(\sim 0.77\) for Power-Law distribution with $\mu = 2.3$, \(\sim 0.67\) for $\mu = 2.5$, and \(\sim 0.59\) for $\mu = 2.7$.}
    \label{fig:DeltavsPerc}
\end{figure}

\noindent
In Table \ref{tab:DFADErelerrfinal} we summarize the best and worst relative errors for $H$ and $\delta$ compared to the corresponding reference errors, while in Figs. \ref{fig:HvsPerc} and \ref{fig:DeltavsPerc} we give a synthetic picture of the mean estimated values of $H$ and $\delta$ with error bars, respectively, compared to their corresponding real values, for all IET distributions, frequency bands and percentiles. From these figures, it is easy to appreciate the better performance of $H$ estimations with respect to the $\delta$ estimations, which display larger fluctuations and greater differences with the theoretical values.
A general remark is the tendency to underestimate the real $H$ and $\delta$ values, even if with some exceptions. This is probably related to the initial short-time transient, which usually displays a slower increase. The automatic best-fitting procedure is able to distinguish the two different regimes, but the short-time slower increase could slightly affect the fit in the long-time regime, where the real scaling exponent emerges.
Interestingly, looking at Table \ref{tab:DFADErelerrfinal} it is clear that the estimation errors (MREs) of $H$ in the power-law cases are comparable to the corresponding reference errors. 
In the best cases, some MREs are surprisingly smaller than the reference errors. Conversely, in the exponential case, the MRE is larger or much larger than the reference error.
%
%%%%%%%%%%%%%%%%%%%%%%%%%%%%%%%%%%%%%%%%%%%%%%%%%%%%%%%%%%%%%%%%%%%%%%%%%%%%%
%%%%%%%%%%%%%%%%%%%%%% Final Tables
%%%%%%%%%%%%%%%%%%%%%%%%%%%%%%%%%%%%%%%%%%%%%%%%%%%%%%%%%%%%%%%%%%%%%%%%%%%%%
% H and delta final tables
\begin{table}[H]
\centering   
    \begin{subtable}[t]{1\linewidth}
    \centering
    \begin{tabular}[t]{c|c|c|c} 
    \multirow{3}{*}{\textbf{Case}} & \textbf{MRE $H$ ($\pm$ STD)} & \textbf{MRE $H$ ($\pm$ STD)} & \textbf{MRE $H$ ($\pm$ STD)} \\[-7pt]
    & \textbf{Reference} & \textbf{Best Case} & \textbf{Worst Case} \\[-7pt]
    & \textbf{Errors} & & \\
      \hline
      \hline
      Exponential & 0.0052 ($\pm$ 0.0042) & 0.010 ($\pm$ 0.010) & 0.038 ($\pm$ 0.017) \\
      \relax
      Power-Law ($\mu = 2.3$) & 0.088 ($\pm$ 0.040) & 0.064 ($\pm$ 0.040) & 0.10 ($\pm$ 0.064) \\
      \relax
      Power-Law ($\mu = 2.5$) & 0.040 ($\pm$ 0.030) & 0.040 ($\pm$ 0.020) & 0.12 ($\pm$ 0.054) \\
      \relax
      Power-Law ($\mu = 2.7$) & 0.070 ($\pm$ 0.040) & 0.018 ($\pm$ 0.013) & 0.071 ($\pm$ 0.021) \\
    \end{tabular}
    \caption{$H$ Relative Errors}
    \label{tab:DFAErrorsRealvsFound}
    \end{subtable}

    \vspace{15pt}
    \begin{subtable}[t]{1\linewidth}
    \centering
    \begin{tabular}[t]{c|c|c|c} 
    \multirow{3}{*}{\textbf{Case}} & \textbf{MRE $\delta$ ($\pm$ STD)} & \textbf{MRE $\delta$ ($\pm$ STD)} & \textbf{MRE $\delta$ ($\pm$ STD)} \\ [-7pt]
    & \textbf{Reference} & \textbf{Best Case} & \textbf{Worst Case} \\ [-7pt] 
    & \textbf{Errors} & & \\
      \hline
      \hline
      Exponential & 0.024 ($\pm$ 0.011) & 0.056 ($\pm$ 0.050) & 0.12 ($\pm$ 0.084) \\
      \relax
      Power-Law ($\mu = 2.3$) & 0.040 ($\pm$ 0.035) & 0.040 ($\pm$ 0.024) & 0.17 ($\pm$ 0.040) \\
      \relax
      Power-Law ($\mu = 2.5$) & 0.062 ($\pm$ 0.021) & 0.17 ($\pm$ 0.018) & 0.28 ($\pm$ 0.065) \\
      \relax
      Power-Law ($\mu = 2.7$) & 0.15 ($\pm$ 0.030) & 0.13 ($\pm$ 0.090) & 0.25 ($\pm$ 0.088) \\
    \end{tabular}
    \caption{$\delta$ Relative Errors}
    \label{tab:DEErrorsRealvsFound}
    \end{subtable}
    \caption{Comparison of MRE for $H$ and $\delta$ estimated from RTEs detected with the RTEF algorithm versus corresponding reference errors, i.e., relative errors of $H$ and $\delta$ estimated from the real RTEs. (a) Reference Errors in $H$ versus the best and worst estimated $H$ from detected RTEs; (b) Reference Errors in $\delta$ versus the best and worst estimated $\delta$ from detected RTEs.}
    \label{tab:DFADErelerrfinal}
\end{table}
%
%%%%%%%%%%%%%%%%%%%%%%%%%%%%%%%%%%%%%%%%%%%%%%%%%%%%%%%%%%%%%%%%%%%%%%%%%%%%%
%%%%%%%%%%%%%%%%%%%%%% Final Plots
%%%%%%%%%%%%%%%%%%%%%%%%%%%%%%%%%%%%%%%%%%%%%%%%%%%%%%%%%%%%%%%%%%%%%%%%%%%%%

% H vs Percentiles

% Delta vs Percentiles
%%%%%%%%%%%%%%%%%%%%%%%%%%%%%%%%%%%
%
\noindent
When looking at the RTE counting,
the best accuracy in RTE detection seems to come
from the higher percentiles, in particular the $98$th one (see Fig. \ref{fig:CountvsPerc}).
Then, we would not expect to find good performances in the evaluation of complexity indices when the number of estimated RTEs is very high concerning the number of real RTEs in the signal. 
Surprisingly, Tables
\ref{tab:PL1DFADEErrors5pt}, \ref{tab:PL2DFADEErrors5pt} and \ref{tab:PL3DFADEErrors5pt} 
show that, for power-law IET distributions, better results for the diffusion scaling exponent $H$ are usually obtained for 
smaller percentiles ($85$, $90$), even if the application of the algorithm results in a very high number of false positives. Conversely, the $98$th percentile, which
better approximates the RTE counts, gives the worst performances in the $H$ evaluation\footnote{
%%%%%%%%%
A slight worsening in accuracy is only seen for
$\mu=2.7$, bands $[4,8]$ and $[8,12]$, when passing from $90$th to $85$th percentiles.
%%%%%%%%%
}.
Actually, power-law cases see an improvement in the estimate of $H$ as the percentile decreases and, therefore, as the number of false positives increases (see, e.g., Tables \ref{tab:PL1DFADEErrors5pt}, \ref{tab:PL2DFADEErrors5pt} and \ref{tab:PL3DFADEErrors5pt}), while
an opposite trend is seen in the exponential case (see Table \ref{tab:ExpDFADEErrors5pt}). In the last case, the $H$ estimation gets worse as the percentile decreases in $[4,8]$ and $[8,12]$ bands and remains about the same in the $[0.5,4]$ band.

\noindent
Conversely, the trend in the $\delta$ evaluation is more intricate, with estimates either getting worse as the percentile decreases or remaining about the same in the power-law case. The only exception we found is given by $\mu=2.7$, where the $\delta$ estimation improves as the percentile decreases.
This last result is similar to the exponential case: $\delta$ evaluation improves as percentile decreases, that is, as the number of false positives increases.
Generally speaking, the exponential case shows the best performances in the evaluation of all parameters, i.e., not only $H$ and $\delta$, but also the direct estimation of $r_p$. This is an expected result, as the exponential distribution has not a strong tail, so that the rare events have a negligible probability, being most of the IET included in 2-3 times the average IET $\langle \tau_p \rangle = 1/r_p$.

\noindent
On the contrary, the opposite trends of both $H$ and $\delta$ estimation accuracy in the exponential case with respect to the inverse power-law distributed IETs are somewhat surprising and are surely related to the very low probability of large IETs of the exponential distribution.
%Conversely, the improvement of $\delta$ scaling with decreasing percentile in the exponential case is somewhat hard to interpret. 
Moreover, we observe that being theoretically $H=\delta=0.5$ in this case, the relative errors are mostly affected by limitations of the fitting procedure.

\noindent
In general, it is important to underline that, in all cases, the mean relative errors in the $\delta$ estimation are much higher than in the $H$ estimation. In fact, being based on the computation of distribution, the DE method needs a greater size of the statistical sample with respect to the DFA method, which essentially evaluates the statistical moments\footnote{
%%%%%%%%%%%%%%%
In this work the second moment, while, in the more general multifractal DFA algorithm \cite{kantelhardt_pa2002-mfdfa}, 
the moments of generic order $q$, including non-integer orders.
%%%%%%%%%%%%%%%
}.
However, as proven by the authors of Ref. \cite{allegrini_pre2002-levywalkbiscaling}, the scaling $\delta$ is more affected by the regions around distribution maxima (the central region for single-mode distributions) and the scaling $H$ is more affected by the distribution tails.
As a consequence, $\delta$ scaling gives different information with respect to scaling $H$ and are both needed in temporal complexity analyses, but the different level of accuracy at the same statistical sample size has to be taken into account.
%has a minimal effect on the estimation of the genuine signal complexity and, in some cases, it seems also to improve the estimation.

\noindent
In a nutshell, the main finding of the present work is that the EDDiS algorithm, which is based on the concept of event-driven diffusion scaling,
is able to decrease the masking effect of false positives and, consequently, the error in the estimation of temporal complexity \cite{grigolini_csf15_bio_temp_complex,paradisi_springer2017}.
This is in agreement with the Time Mixed Model (TMM), where a mixing of complex events and Poisson noisy events is investigated \cite{paradisi_csf15_pandora,allegrini_pre10}.
However, these studies were limited to the case of a low rate of noisy events with respect to that of complex events.
Here, the noisy events are given by the numerous false positives in the event-detection algorithm and, in some case, the rate of noisy events is much larger than that of the genuine complex events.
This shortcoming of the event detection algorithm was here found not to compromise the possibility for the EDDiS algorithm of obtaining a good estimate of complexity, particularly of the scale of the second $H$ moment, in the power-law case.
%Perch\'e nell'exponential case si somma un altro Poisson, con un altro event rate, ma che darebbe circa sempre $0.5$, ma con un lieve peggioramento. Invece, nel caso power-law, si sovrappone un Poisson, che viene compresso ai tempi iniziali, quindi potrebbe migliorare il fit.
In fact, regarding the application of DFA, the main, but counter-intuitive, result is a net improvement in the performances of complexity estimation, in the power-law case, as the number of false positives in the RTEF algorithm increases. Conversely, the trend is the opposite for the exponential case, as it would be expected for all cases.

% Iniziare con un breve recap delle motivazioni ed obiettivi che hanno potato alla realizzazione di questo algoritmo

% Fare un riassunto di tutti I segnali creati per testare questo algoritmo

% Fare un riassunto delle analisi fatte per quantificare le performance di questo algoritmo

% Parlare dei principali risultati ottenuti con questo algoritmo, facendone anche un riassunto dei vantaggi nell'uso di questo algoritmo ma sottolineando I limiti e I possibili futuri miglioramenti

\section*{Acknowledgements}

\noindent
This work was supported by the Next-Generation-EU programme under the funding schemes PNRR-PE-AI scheme (M4C2, investment 1.3, line on AI)
FAIR “Future Artificial Intelligence Research”, grant id PE00000013, Spoke-8: Pervasive AI.

\section*{Authors' contribution}

\noindent
Authors contributed equally to this work.



%%       \citet{<label>} ==> Jones et al. [21]
%%       \citep{<label>} ==> [21]
%%

\bibliographystyle{elsarticle-num-names} 
\bibliography{Bibliography}


%     \bibitem[Authors et al. (Year)]{<label>}
% \end{thebibliography}

\end{document}